\documentclass[12pt,reprint,aps,prb,floatfix]{revtex4-1}
\usepackage{epsfig}
\usepackage{amsmath}
\usepackage[dvipsnames]{xcolor}
\usepackage{amsfonts}
\usepackage{amssymb}
\usepackage{booktabs}
\usepackage{multirow}
\usepackage{soul}
\usepackage{graphicx}
\usepackage{tabularx}
\usepackage{siunitx}
\usepackage{colortbl} 
\usepackage[version=3]{mhchem}
\usepackage{longtable}
\usepackage[colorlinks=true, allcolors=blue]{hyperref}

\begin{document}

\title{Interpretable Machine Learning for Thermoelectric Materials Design with Kolmogorov--Arnold Networks}

\author{Marco Fronzi}
\email{marco.fronzi@sydney.edu.au}
\affiliation{School of Physics, The University of Sydney, Camperdown, Australia}

\author{Michael J Ford}
\affiliation{School of Mathematics and Physics, The University of Technology Sydney, Haymarket, Australia}

\author{Kamal Singh Nayal}
\affiliation{Department of Chemistry, Carnegie Mellon University, Pittsburgh, United States}

\author{Olexandr Isayev}
\affiliation{Department of Chemistry, Carnegie Mellon University, Pittsburgh, United States}

\author{Catherine Stampfl}
\affiliation{School of Physics, The University of Sydney, Camperdown, Australia}

\date{\today}

\begin{abstract}
The discovery of high-performance thermoelectric materials requires models that are both accurate and interpretable. 
Traditional machine learning approaches, while effective at property prediction, often act as black boxes and provide 
limited physical insight. In this work, we introduce Kolmogorov--Arnold Networks (KANs) for the prediction of 
thermoelectric properties, focusing on the Seebeck coefficient and band gap. Compared to multilayer perceptrons (MLPs), KANs achieve comparable predictive accuracy while offering explicit symbolic representations of structure-property relationships. This dual capability enables both reliable predictions and physically interpretable functional forms, providing insight into the governing mechanisms of thermoelectric behaviour. Benchmarking against literature baselines highlights their robustness and generalisability, demonstrating that KANs constitute a practical framework for reverse engineering materials with targeted thermoelectric performance and bridging the gap between predictive power and scientific interpretability.
\end{abstract}
\maketitle

\section{Introduction}

Thermoelectric materials show significant promise for a variety of applications ranging from power generation to refrigeration.\cite{Bell2008} Recent reviews have reaffirmed their potential for energy sustainability and waste heat recovery.\cite{Yan2022} The efficiency of these materials is fundamentally tied to their Seebeck coefficient, electrical conductivity, and thermal conductivity.\cite{Snyder2008} However, the discovery and design of materials with efficient energy conversion remains a substantial challenge in the field.\cite{Snyder2008,Beretta2019,Liu2015} A key metric used to assess thermoelectric performance is the dimensionless figure of merit, $zT$, defined as:

\begin{equation}
zT = \frac{S^2 \sigma T}{\kappa},
\end{equation}

where $S$ is the Seebeck coefficient, $\sigma$ the electrical conductivity, $T$ the absolute temperature, and $\kappa$ the total thermal conductivity.\cite{Snyder2008} The main difficulty lies in the complex interdependence between thermal and electrical conductivity, which cannot be easily decoupled to achieve a high $zT$.\cite{Wolf2019,Su2022}

To understand these intricate relationships, quantum mechanical calculations --particularly those based on density functional theory (DFT) -- can be employed.\cite{Gorai2017,Gutierrez2020} DFT provides invaluable insights into the electronic structure and transport properties of materials, thereby guiding the design and discovery of new thermoelectric compounds.\cite{Kohn1965,Gorai2017,Pohls2021} However, such calculations are often computationally intensive and time-consuming, limiting their scalability for high-throughput screening.\cite{Curtarolo2013}

Machine learning (ML) has emerged as a transformative tool with the potential to revolutionise materials discovery.\cite{Butler2018} When trained on existing data from quantum mechanical simulations and/or experimental results, ML models can rapidly predict the properties of novel materials, dramatically accelerating the pace compared to traditional computational methods.\cite{Raccuglia2016} Furthermore, these models can capture complex, non-linear patterns in data and generate accurate predictions across large datasets, making them particularly well-suited for exploring vast materials spaces.\cite{Xie2018} However, a major limitation of most ML approaches is their reliance on correlation rather than causation, which often prevent the understanding of the underlying physical mechanisms.\cite{Oviedo2022,Zhong2022} As a result, their predictions, although highly valuable, typically serve as guideline rather than definitive explanations in the materials discovery process.\cite{Oviedo2022,Zhong2022,SchmidtHieber2021}
These limitations motivate the exploration of interpretable machine learning architectures capable of capturing the underlying physics rather than merely correlating descriptors with target properties.

The Kolmogorov-Arnold Network (KAN) represents a modern neural network design inspired by the Kolmogorov-Arnold representation theorem, which demonstrates that any multivariate continuous function can be expressed as a sum of one-dimensional functions of linear combinations of the inputs.\cite{SchmidtHieber2021,Liu2025} The KAN implements this concept through learnable univariate spline-based activation functions arranged in layers followed by summation, in contrast to the fixed nonlinearities used in traditional neural networks.\cite{Liu2025} 
This design not only ensures universal approximation capability--allowing the representation of any continuous function--but also enforces local smoothness and continuity of the learned mapping. 
These properties later translate into enhanced stability and physical interpretability when modelling complex, correlated electronic systems.
%
 Kolmogorov-Arnold Networks allow better interpretability because each hidden unit functions as a one-dimensional relationship between a particular linear combination of input features and the target property. This unique architecture, coupled with learnable activation functions, provides a direct functional decomposition: the target property acquires an analytic functional form, which is highly useful for thermoelectric modelling due to the complex, coupled nature of the Seebeck coefficient and thermal conductivity. KANs provide explicit functional decomposition to disclose physical connections in thermoelectric materials, which conventional black-box models may hide, establishing an analytically sound method to model multivariable functions and enabling meaningful interpretation of predictions.\cite{He2015} The Kolmogorov-Arnold framework establishes an analytically sound method to model multivariable functions, which enables meaningful interpretation of predictions.

While KANs are promising for capturing such physics-informed relationships, their performance must be contextualised against widely adopted black-box models to assess trade-offs in interpretability and accuracy.

In this study, we present an integrated machine learning framework for predicting the Seebeck coefficient--a key descriptor of thermoelectric performance--across a diverse set of bulk materials. 
To further evaluate the versatility of our approach beyond transport properties, we extend the modelling framework to predict the electronic band gap---an equally important but physically distinct property that governs many aspects of thermoelectric behaviour. Although not a direct transport coefficient, the band gap is a fundamental electronic property that strongly influences thermoelectric performance by affecting intrinsic carrier concentration, electrical conductivity, and bipolar conduction, particularly at elevated temperatures.\cite{chen2021expand} Its inclusion serves a dual purpose: first, as a complementary screening metric that reflects the quality of the underlying electronic structure; and second, as a benchmark for assessing the flexibility and generalisability of the KAN architecture. Unlike the Seebeck coefficient, which is highly sensitive to the curvature and asymmetry of the bands near the Fermi level,\cite{graziosi2022bipolar,qiu2025revealing} the band gap is governed by broader features of the electronic structure. Strong performance across both properties demonstrates the model’s robustness in capturing physical trends that span from global electronic characteristics to fine-grained transport behaviour.

To provide a rigorous benchmark, we also train a multi-layer perceptron (MLP) on the same dataset, enabling direct comparison between the interpretability-accuracy trade-offs of KAN and a conventional deep learning architecture.

\section{Methodology}

The methodological framework of this study combines conventional machine learning baselines with novel interpretable neural architectures to predict key electronic and thermoelectric properties of crystalline materials. 
Our objective is twofold: first, to establish a reliable reference using well-understood models, and second, to assess the performance and interpretability gains enabled by Kolmogorov--Arnold Networks. 
To this end, we employed multilayer perceptrons (MLPs) as benchmark models, providing a standard against which KAN results can be rigorously compared.

\subsection*{Multilayer Perceptrons as Benchmark Models}  

\begin{figure}[htbp]
    \centering
    \includegraphics[width=0.33\textwidth]{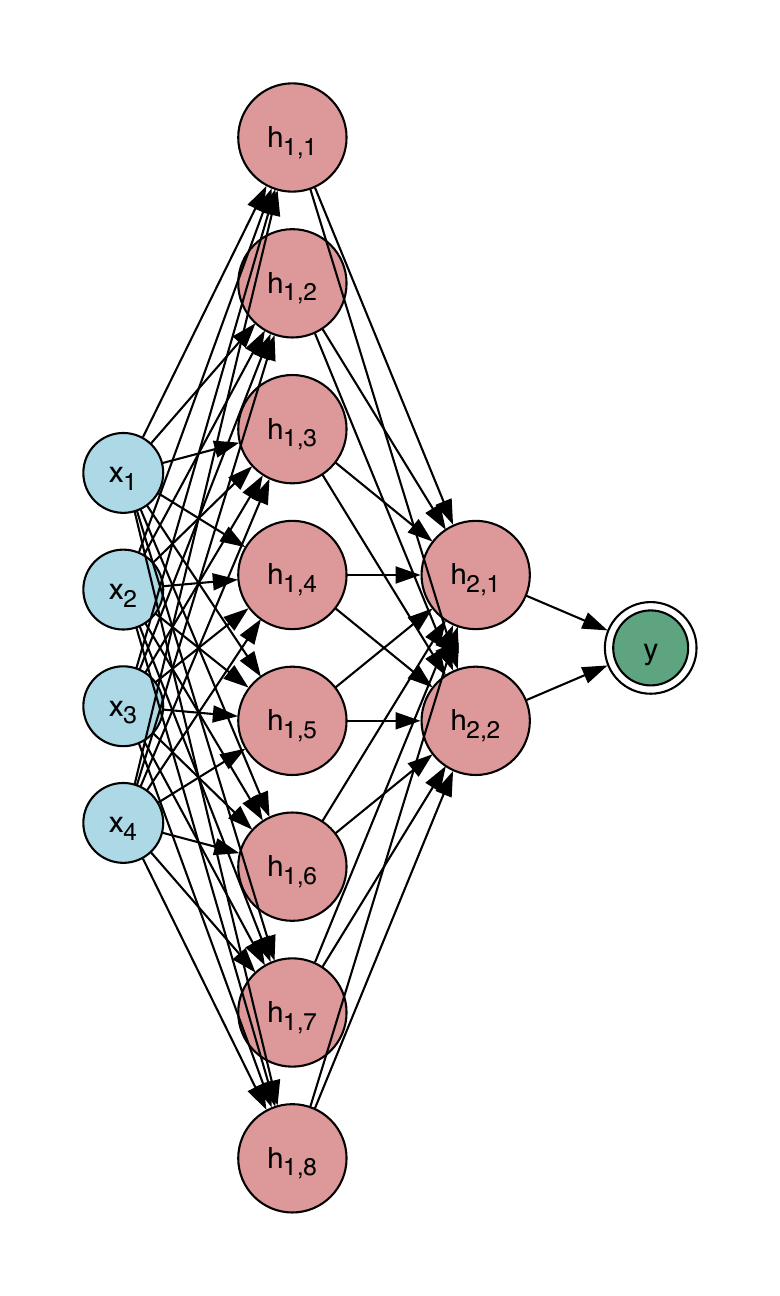} 
    \caption{Schematic of a standard feedforward neural network. Each hidden node computes a weighted sum of its inputs, applies a non-linear activation function, and propagates the signal forward. The output is a scalar regression target, such as the Seebeck coefficient or band gap.}
    \label{fig:NN-graph}
\end{figure}

An MLP is a fully connected feedforward neural network in which nodes are arranged in successive layers, as shown in Fig~\ref{fig:NN-graph}. Each neuron computes a weighted sum of its inputs followed by a nonlinear activation:  
\begin{equation}
f(x_1, \ldots, x_n) = \sigma\left( \sum_{i=1}^{n} w_i x_i + b \right),
\end{equation}
where \(w_i\) and \(b\) denote the learnable weight and bias parameters, and \(\sigma\) represents the activation function. In this work, rectified linear units (ReLU) were chosen for the hidden layers owing to their computational efficiency and ability to mitigate vanishing gradients, while a linear activation was adopted in the output layer to accommodate regression tasks such as band gap and Seebeck coefficient prediction.  

The models were implemented in the \texttt{PyTorch} framework and trained on datasets split into training (80\%) and test (20\%) subsets, with performance further validated using 5-fold cross-validation, and early stopping was triggered with a patience of 80 epochs based on validation loss.  

Hyperparameters were selected through systematic grid search across layer sizes, learning rates, and patience values. The Adam optimiser was employed with a fixed learning rate of \(10^{-3}\), weight decay of \(10^{-4}\), and default momentum parameters \((\beta_1 = 0.9, \beta_2 = 0.999)\).\cite{paszke2019pytorch} Training minimised the mean squared error (MSE) loss function and proceeded for up to 2000 epochs. Model selection was guided by multiple metrics, including the coefficient of determination (\(R^2\)), root mean squared error (RMSE), and mean absolute error (MAE), ensuring accurate predictions with minimal train--test degradation.  

The optimal architecture identified through this process was \([128, 64, 4, 1]\). This configuration consistently achieved high \(R^2\) values alongside low RMSE and MAE across validation folds. Importantly, the small discrepancies between training and test metrics highlighted the strong generalisation ability of the model, validating its role as a robust benchmark against which KAN performance could be assessed.  

\subsection{Kolmogorov--Arnold Networks}  

Kolmogorov--Arnold Networks are a neural architecture, schematically shown in Fig.~\ref{fig:KAN-graph} derived from the Kolmogorov--Arnold representation theorem, which guarantees that any multivariate continuous function can be expressed as a finite superposition of univariate continuous functions combined with binary addition operations.\cite{kan2023,liu2024kan,schmidt2021kolmogorov}  
Formally, the representation is written as:
\begin{equation}
\label{eq:kan}
f(x_1, \ldots, x_n) = \sum_{q=0}^{2n} \Phi_q \left( \sum_{p=1}^{n} \phi_{q,p}(x_p) \right),
\end{equation}
\noindent


The index $q$ runs from $0$ to $2n$, ensuring a finite set of superpositions. In this representation, $\phi_{q,p}$ denotes the inner univariate function acting on input $x_p$, and $\Phi_q$ denotes the outer univariate function applied to the aggregated contributions from all inputs. This decomposition establishes the theoretical foundation for universal approximation when implemented with shallow, fixed-width networks, which require at most $2n+1$ hidden nodes.~\cite{kan2023}

 Unlike multilayer perceptrons (MLPs), which place fixed nonlinear activations at nodes and linear transformations on edges, KANs invert this design:  
\begin{itemize}
    \item \textbf{Nodes} act as summation units.  
    \item \textbf{Edges} implement learnable nonlinearities parameterised as one-dimensional spline functions.  
\end{itemize}

This edge-based parametrisation provides intrinsic interpretability, as each connection corresponds to an explicit functional transformation of its input. Consequently, KANs are particularly well suited for tasks such as symbolic regression and descriptor discovery. 
We reconfigure the model to evaluate symbolic representation, where the output of each neuron in a KAN layer can be expressed as:

\begin{equation}
\label{eq:kan_layer}
    y_i = \sum_{j} w_{ij} \, \phi_{ij}(x_j),
\end{equation}

where $x_j$ denotes the $j$-th input feature, $\phi_{ij}(\cdot)$ is a trainable univariate function associated with the connection from input $j$ to neuron $i$, and $w_{ij}$ is a scalar weight. This formulation can be seen as the practical, layer--wise realisation of the Kolmogorov--Arnold functional decomposition shown in Equation~(\ref{eq:kan}), in which each multivariate function is represented as a finite sum of outer functions $\Phi_q$ applied to inner sums of univariate transforms $\phi_{q,p}$. In the network implementation, the inner sum over $p$ corresponds to the aggregation $\sum_j w_{ij} \, \phi_{ij}(x_j)$, while the outer function $\Phi_q$ is either absorbed into the next layer or represented by subsequent $\phi$ transformations. This mapping bridges the theorem-level representation and the computational architecture, preserving the universal approximation property while enabling gradient-based optimisation. 
 
Here, KANs were implemented using the open-source \texttt{PyTorch} and \texttt{pykan} libraries, which provide full hyperparameter control and differentiable optimisation of spline-based activations.\cite{paszke2017automatic,liu2024kan} Each spline was initialised on a fixed grid and parameterised by cubic B-splines ($k=3$), with grid resolution $G=12$. The learning objective combined standard prediction error minimisation with $\ell_1$-type sparsity regularisation, promoting compact models. 


The network architecture consisted of an input layer matching the feature dimension (128), one hidden layer (width 16), and a single output node. Training employed the Adam optimiser in the initial phase, whereas the final model refinement was performed using the limited-memory Broyden--Fletcher--Goldfarb--Shanno (LBFGS) optimiser.  Learning rate and weight decay were set to $10^{-3}$ and $10^{-4}$, respectively, for up to 2000 epochs, with early stopping (patience of 80 epochs) based on validation MSE. Two additional regularisation terms were incorporated: an $\ell_1$ sparsity penalty ($\lambda = 0.01$) to promote compactness, and an entropy-based smoothness penalty ($\lambda_{\text{entropy}} = 0.1{-}0.2$) to stabilise functional representations.
The regularisation term is weighted by $\lambda_{\text{entropy}}$, which controls the smoothness of the learned spline functions. 
In the Gaussian Mixture Model (GMM) used for output scaling, $\alpha$ is the mixing coefficient determining the relative contribution of each Gaussian component, and $p_\mathrm{GMM}$ represents the corresponding membership probability.

\subsubsection*{Symbolic extraction and interpretability}
After the numerical optimisation, symbolic post-processing was performed to reconstruct analytic surrogates for the most influential spline functions.
 Models clarity was improved by pruning weak or non-contributing edges, identified by coefficient magnitudes below a fixed threshold, set to 0.01. Removing these edges reduced complexity and eliminated noisy contributions, yielding more compact and physically meaningful symbolic expressions.

The symbolic form of each learned activation was then extracted by fitting candidate functions from a predefined library ($\mathcal{D}$), which in its final form included a broad set of elementary functions (Fig.~\ref{list:basis_functions}):

\begin{figure}[htbp]
    \centering
    \includegraphics[width=0.4\textwidth]{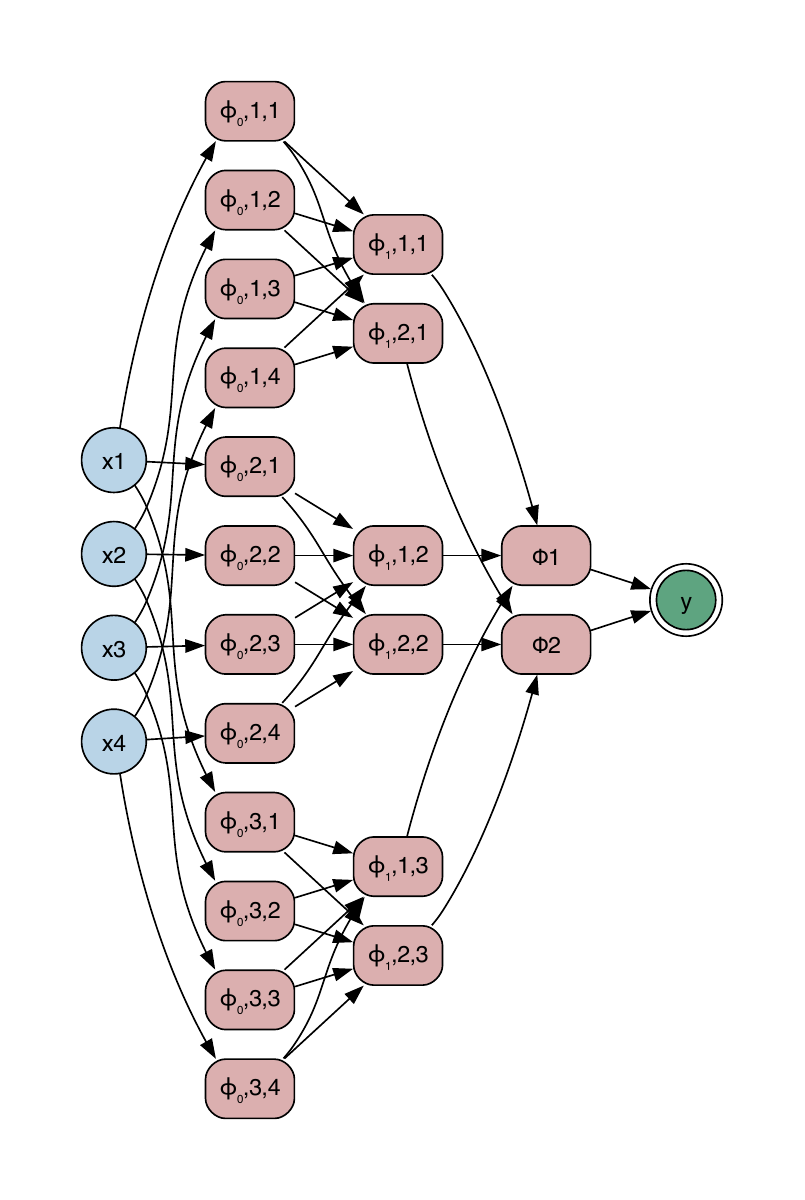} 
    \caption{Schematic representation of a Kolmogorov--Arnold Network. Each input $x_i$ is transformed by a set of learnable univariate functions $\phi_{q,p}(x_p)$, summed, and then mapped by outer functions $\Phi_q$ to produce the final output $y$. This architecture directly embodies the compositional structure dictated by the Kolmogorov--Arnold representation.}
    \label{fig:KAN-graph}
\end{figure}

\begin{figure}[h!]
\centering
\begin{minipage}{0.8\textwidth}
\begin{itemize}
  \item $x$, $x^2$, $x^3$, $x^4$, $x^5$
  \item $\frac{1}{x}$, $\frac{1}{x^2}$, $\frac{1}{x^3}$, $\frac{1}{x^4}$, $\frac{1}{x^5}$
  \item $\sqrt{|x|}$, $x^{0.5}$, $x^{1.5}$, $\frac{1}{\sqrt{|x|}}$, $\frac{1}{x^{0.5}}$
  \item $e^{x}$, $\log(|x|+\epsilon)$, $|x|$
  \item $\sin(x)$, $\cos(x)$, $\tan(x)$, $\tanh(x)$
  \item $\text{sgn}(x)$, $\arcsin(x)$, $\arccos(x)$, $\arctan(x)$, $\text{arctanh}(x)$
  \item $0$
  \item $e^{-x^2}$
\end{itemize}
\end{minipage}
\caption{ Set of elementary functions used as basis candidates ($\mathcal{D}$).}
\label{list:basis_functions}
\end{figure}

where $\epsilon$ is a small positive constant (e.g., $10^{-6}$) added to avoid singularities in logarithmic and reciprocal functions.

This broad set allowed for flexible representation of both polynomial and non-polynomial relationships, oscillatory patterns, and asymptotic behaviours observed in thermoelectric descriptors.

Each active univariate spline $\phi_{ij}(x)$ is approximated with a compact symbolic surrogate $\tilde{\phi}_{ij}(x)$ drawn from a dictionary $\mathcal{D}$. 
 Candidate expressions are fitted by least squares on the knot grid and validated on held-out points sampled within the empirical support of $x$. 
%
The reported complexity score \(c\) in the tables is a discrete proxy for interpretability:
\begin{itemize}
    \item \(c=1\): affine or single low-order polynomial ($x$, $x^2$).
    \item \(c=2\): single-elementary nonlinearity with bounded range or simple rational form (e.g., $\sin$, $\cos$, $1/x$) with one affine phase/scale.
    \item \(c=3\): higher-curvature or singular forms (e.g., $\tan$) or shallow compositions of two primitives (e.g., $\sin(ax+\beta)+\gamma x$).
    \item \(c\ge 4\): piecewise or multi-term compositions (not used when a lower-$c$ surrogate attains $r^2\ge 0.996$).
\end{itemize}

To extract symbolic relations from the trained models, we employed an 
$R^2$ acceptance threshold for symbolic regression. We set the cutoff 
value to $R^2 = 0.9$ provided the 
best trade-off between interpretability and reliability. At this threshold, the 
symbolic functions retained sufficient accuracy to capture the dominant 
structure--property relationships, while still allowing the inclusion of 
approximate functional forms that may reflect underlying physical trends. 

The extracted analytical expressions facilitated the mapping of learned relationships back to physically interpretable descriptors, enabling direct comparison with known theoretical forms and empirical trends.

\subsection{Dataset and Target Properties}

The dataset used in this study is derived from the \texttt{ricci\_boltztrap\_mp\_dataset}, which contains thermoelectric and electronic properties computed via DFT followed by semi-classical Boltzmann transport analysis using the BoltzTraP code.\cite{ricci2017ab,ricci2018figshare,madsen2006boltztrap} The crystal structures originate from the Materials Project database, ensuring consistent treatment of exchange--correlation effects and structural optimisation parameters across the dataset.\cite{jain2013materials_project}

The target properties include the electronic band gap and the Seebeck coefficients for electrons (\(S_n\)). Band gaps were obtained from standard DFT calculations and subsequently corrected to improve alignment with experimental trends. The Seebeck coefficients were calculated under the constant relaxation time approximation by solving the linearised Boltzmann transport equation, using a fixed temperature of 300~K and a chemical potential aligned with the intrinsic Fermi level. Units are \(\mu\mathrm{V/K}\) and \(S_n\), and electronvolts (eV) for the band gap.

Although these computed quantities are widely used in materials informatics, they are subject to well-known limitations. For band gaps, standard DFT tends to underestimate absolute values due to the lack of quasi-particle corrections. For Seebeck coefficients, the constant relaxation time approximation neglects scattering mechanism variations, and the assumption of a fixed temperature ignores potential thermal dependencies. Furthermore, BoltzTraP calculations assume parabolic band shapes near the Fermi level, which can introduce systematic deviations for materials with highly non-parabolic dispersions. Despite these factors, the relative trends and rank ordering of materials are generally preserved, and such limitations are negligible in the present work, as the primary objective is the evaluation of KAN architectures for predicting complex target properties and interpreting the learned structure--property relationships, assessing the utility of the models for reverse engineering purposes.

\begin{figure}[htbp]
    \centering
    \includegraphics[width=0.4\textwidth]{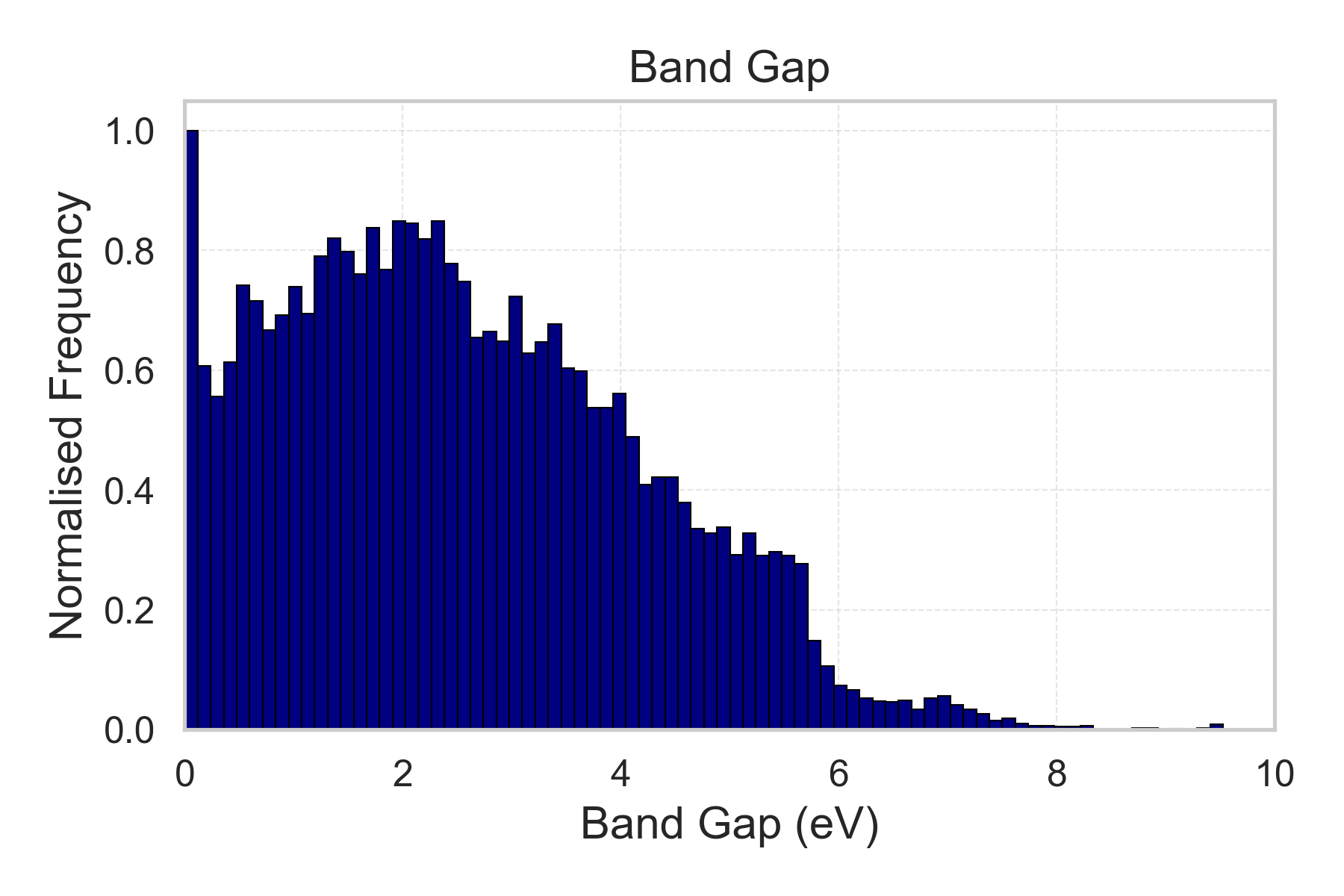} \\
    \includegraphics[width=0.4\textwidth]{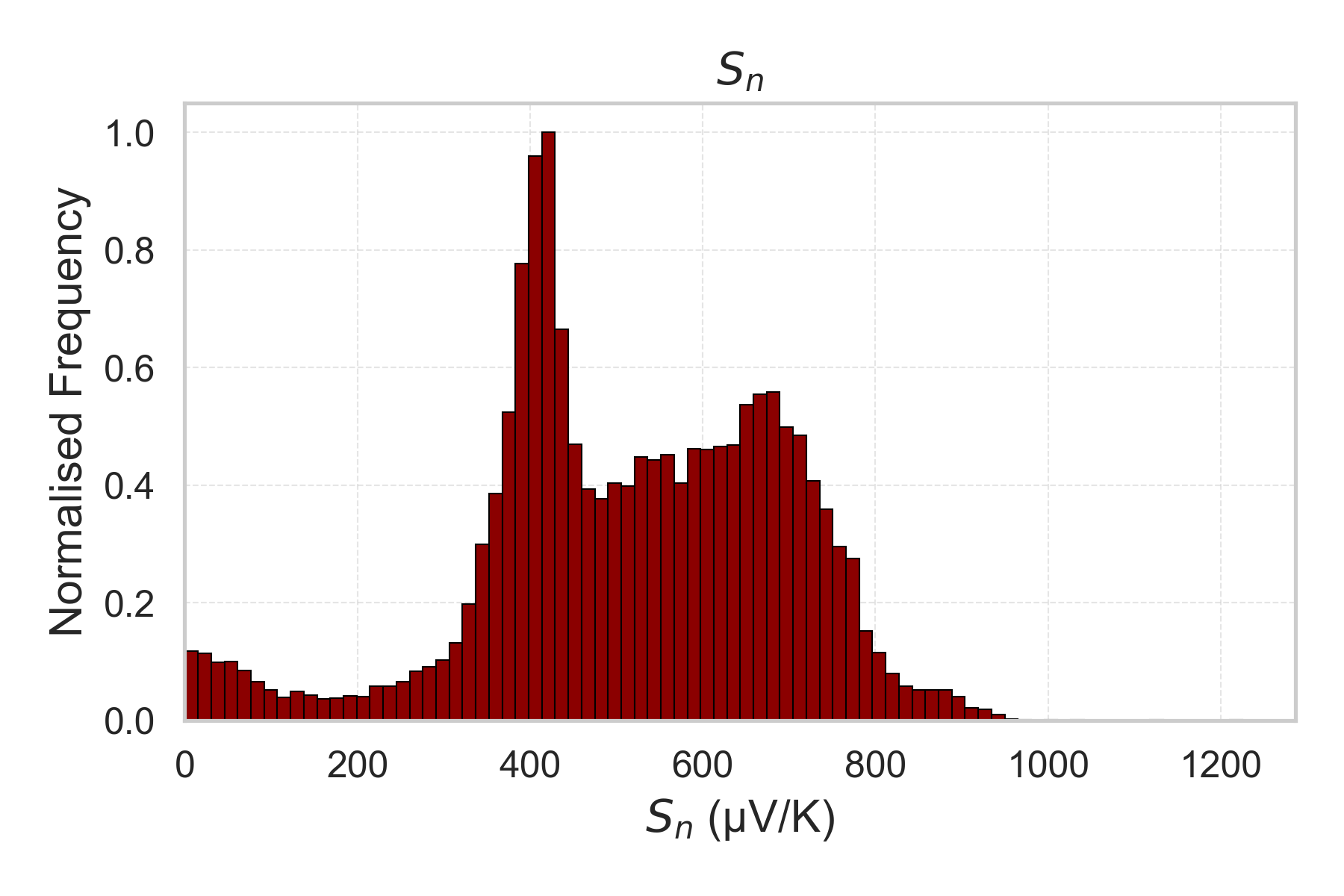} \\
    \caption{Histograms showing the distribution of key target properties in the dataset.
        \textbf{Top:} Band gap, which is skewed right with a large number of materials having small band gaps and a long tail extending beyond 6~eV.
        \textbf{Bottom:} Seebeck coefficient for electrons (\( S_n \)), displaying a bimodal distribution with peaks near 400~\(\mu\)V/K and 650~\(\mu\)V/K.}
    \label{fig:target_histograms}
\end{figure}

The statistical distribution of the target properties is shown in Fig.~\ref{fig:target_histograms}. The band gap distribution is right-skewed, with a large proportion of materials exhibiting small band gaps and a long tail extending beyond 6~eV, whereas the electron Seebeck coefficient exhibits a left-skewed bimodal distribution with peaks near 400~\(\mu\)V/K and 650~\(\mu\)V/K.

\subsection{Crystalformer Representations and Feature Encoding}

To describe the crystal structure space effectively, we use \textit{Crystalformer}, a novel Transformer framework for periodic structure encoding. Crystalformer is a Transformer-based encoder inspired by Graphormer, which applies fully connected attention between atoms in a molecule.\cite{taniai2024Crystalformer,ying2021transformers} To capture both local atomic environments and long-range interactions in particular, it extends this approach to introduce infinitely connected attention arising from the periodicity in crystals, formulated as an infinite summation of interatomic potentials in an abstract feature space. The Crystalformer architecture follows the Transformer\cite{vaswani2017attention} encoder design with stacked self-attention blocks made of two residual connections linking a multi-head attention layer and a shallow feed-forward network, but unlike the original model it removes Layer Normalization entirely to help stabilize training.\cite{vaswani2017attention} It builds upon the success of materials graph networks while incorporating positional (both spatial and edge) encoding for periodicity-aware modeling and ensuring permutation, SE(3) and periodic invariance (both supercell and periodic-boundary shift).

The use of Crystalformer embeddings provides several advantages: (i) it captures high-order geometric correlations and structural motifs not easily represented by classical hand-crafted features; (ii) it enables the transfer of knowledge from large crystal datasets to our thermoelectric prediction task; and (iii) it supports end-to-end differentiability and integration with downstream prediction models. The physics-inspired treatment of infinitely connected attention leads to learned structural embeddings, which enhance both the predictive accuracy and robustness of the models trained on them.

In this work, Crystalformer was used as a \textit{supervised featuriser} that transforms input crystal structures into fixed-length, continuous vector embeddings. Each structure, initially represented by its atomic positions, species, and lattice parameters, was encoded into a learned structural representation by training on a large dataset of materials from the Materials Project and the Open Quantum Materials Database.\cite{saal2013oqmd,jain2013materials_project} Starting with a set of trainable atom embeddings representing the atomic species of the unit cell as the initial state, Crystalformer transforms them into an abstract state through four stacked self-attention blocks using neural potential summation for capturing crystal periodicity. The atom-wise states in the abstract state are then aggregated into a single vector via global average pooling, which serves as the latent embedding.

To integrate these latent representations into our predictive pipeline, we extracted the final vector embeddings from the penultimate layer of the Crystalformer encoder, resulting in a 128-dimensional vector for each input material. Prior to model training, all features were standardised to zero mean and unit variance using the training set statistics. Crystalformer code and training settings were directly adapted from the original paper and modified to account for multi-target training, as well as to save the final layer representation before the regression layer for use as the latent representation of the input crystal structures. All training data was used to train the Crystalformer model with a batch size of 256 materials for 250 epochs.

\subsubsection*{KAN attribution scores}

To assess the relevance of individual descriptors we used the built-in attribution analysis available in the Kolmogorov--Arnold Network framework. 
In KANs, each edge between nodes carries an adaptive spline function that directly maps input values to activations. 
During attribution, the network is first evaluated on a representative dataset to record the activations of all spline functions. 
The contribution of each input descriptor to the final output is then quantified by aggregating the absolute magnitudes of the learned spline functions along all paths that connect the input to the output node. 
Formally, the attribution score for descriptor $x_i$ is defined as
\begin{equation}
\label{eq:kan_attr}
S_i = \frac{1}{Z}\,\sum_{p \in \mathcal{P}(i \to y)} 
\prod_{(u \to v) \in p} \big| f_{uv}(a_u) \big|,
\end{equation}
where $\mathcal{P}(i \to y)$ denotes the set of all directed paths from input $x_i$ to the output node $y$, 
$f_{uv}$ is the spline function along edge $(u \to v)$, $a_u$ is the activation of node $u$, and $Z$ is a normalisation factor ensuring $\sum_i S_i = 1$. 
This procedure yields a feature attribution score for every descriptor, which can be interpreted as a normalised measure of its overall influence on the target prediction. 
Unlike gradient-based sensitivities, which reflect local responsiveness of the output to infinitesimal perturbations, 
KAN attribution scores incorporate the full functional form of the spline edges, and thus provide a more global estimate of descriptor importance consistent with the symbolic structure of the trained model.

\section{Results}

\subsection{Data Preprocessing and Target Normalisation}

\subsubsection*{Feature scaling}
The input feature matrix \(\mathbf{X}\), comprising structural and compositional descriptors extracted from learned embeddings, was standardised to zero mean and unit variance using \texttt{StandardScaler}:
\begin{equation}
\mathbf{X}_{\text{scaled}} = \frac{\mathbf{X} - \mu}{\sigma},
\end{equation}
\noindent
where \(\mu\) and \(\sigma\) denote the column-wise mean and standard deviation, respectively. Standardisation mitigates internal covariate shift and accelerates convergence in neural architectures such as MLPs and KANs.

\subsubsection*{Target scaling}
Because the target variables exhibited markedly different statistical distributions, 
property-specific normalisation procedures were applied to improve numerical stability and regression accuracy. 
Each property---the electronic band gap and the electron Seebeck coefficient ($S_n$)---was modelled independently. 
Preprocessing was carried out in \texttt{Python} using the \texttt{scikit-learn} library, and the fitted scalers were stored with \texttt{joblib} to guarantee consistent transformations during training and inference. 
The normalisation was designed to approximate Gaussian-like target distributions, a choice that facilitates convergence and stabilises optimisation in gradient-based learning algorithms.

\paragraph{Band gap.}
The DFT-predicted band gaps were approximately unimodal and symmetric; thus, a standard \(z\)-score transformation using \texttt{StandardScaler} was applied:
\begin{equation}
y_{\text{scaled}}^{(\text{band\_gap})} = \frac{y - \bar{y}}{\mathrm{std}(y)}.
\end{equation}

\paragraph{Electron Seebeck coefficient (\(S_n\)).}

For the Seebeck coefficients, strongly non-Gaussian, bimodal and strongly skewed  distributions required composite normalisation procedures.

To address this, a multi-step transformation was employed:
(i) log-sign transformation to suppress heavy tails:
\begin{equation}
y' = \text{sgn}(y) \cdot \log(1 + |y|),
\end{equation}
(ii) followed by Gaussian Mixture Model (GMM) normalization to approximate a Gaussian-like target distribution suitable for gradient-based optimization.
\begin{equation}
y_{\text{scaled}}^{(S_n)} = \mathrm{GMM\_Scaler}(y'),
\end{equation}
where $\mathrm{GMM\_Scaler}$ is a composite transformation using the learned parameters of the Gaussian components, a necessity for robustly modeling the \textbf{bimodal distribution} observed for the electron Seebeck coefficient.

\subsubsection*{Rationale}
These transformations were designed to preserve the physical interpretability of the target variables while improving their suitability for learning with smooth, spline-based architectures. By addressing skewness, heavy tails, and multimodality, the preprocessing pipeline enhances both model convergence and predictive stability.

\subsection{Multilayer Perceptron Network Baseline Performance}

\begin{figure}[htbp]
  \centering
  \includegraphics[width=0.5\textwidth]{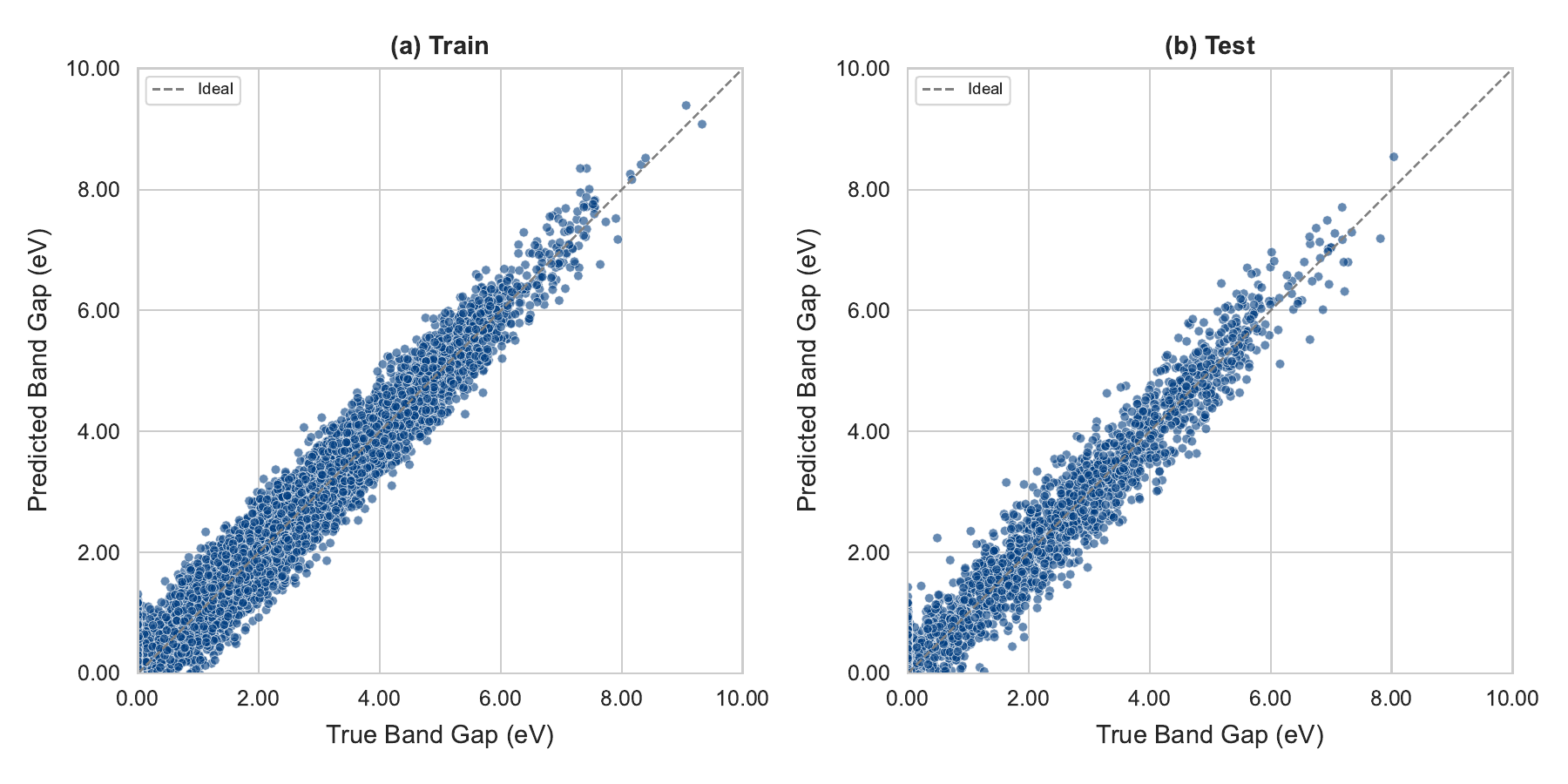} \\
  \includegraphics[width=0.5\textwidth]{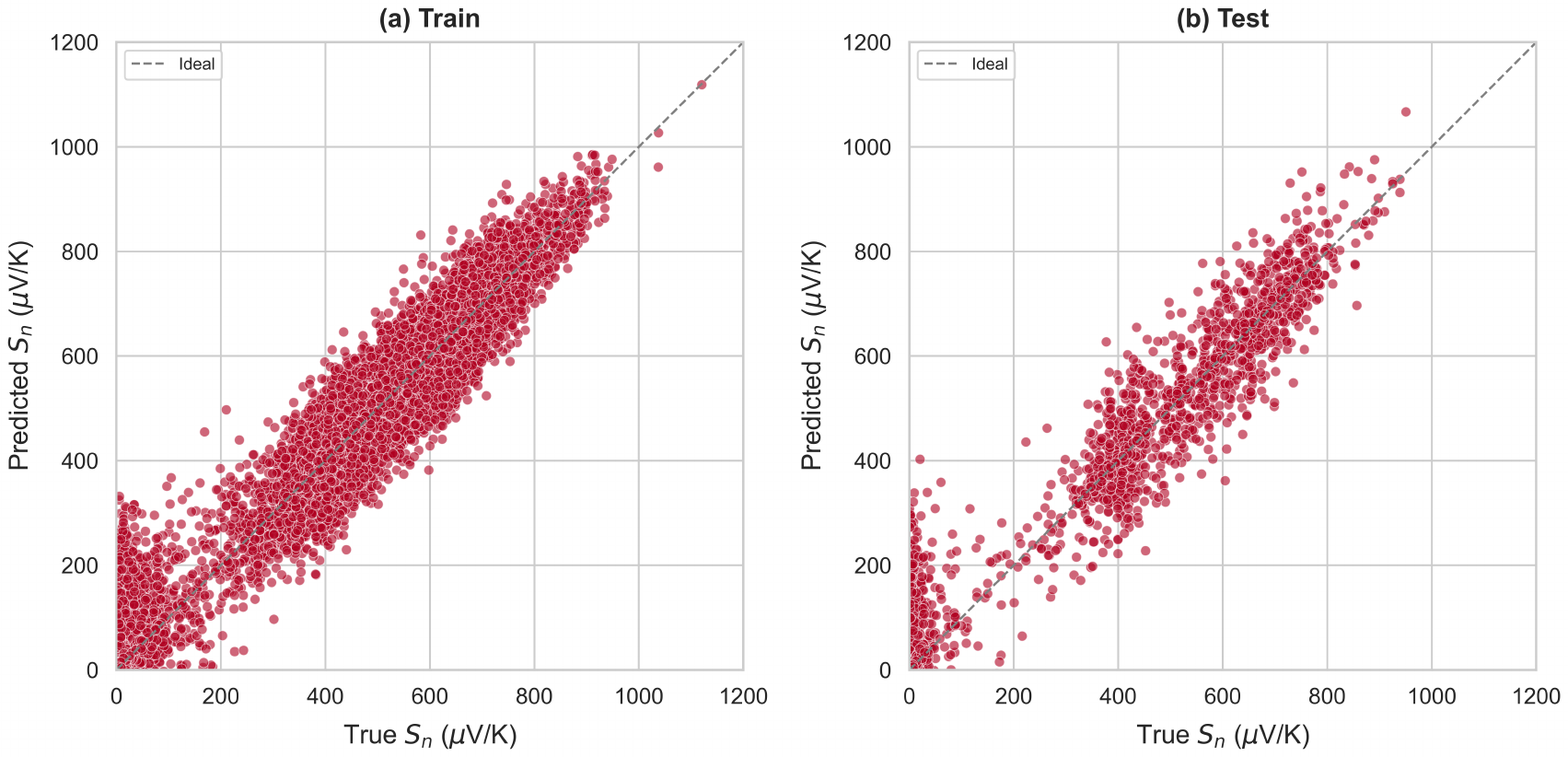} \\
  \caption{Parity plots for predicted vs. true values of the band gap and $S_n$ (top and bottom, respectively) using the MLP model. Each property is shown for both training and test sets (left to right). The diagonal line represents perfect prediction.}
  \label{fig:mlp-parity-all}
\end{figure}

We implemented a fully connected neural network MLP baseline with architecture $[128,64,4,1]$, selected via grid search on CrystalFormer embeddings using $R^2$ and MSE as reference metrics, and trained it to predict the band gap and Seebeck coefficient. The filtered dataset includes 15{,}000 structures split 80/20 into training and test sets. Figure~\ref{fig:mlp-parity-all} presents parity plots for train and test partitions, while Table~\ref{tab:mlp_metrics} reports the corresponding metrics. 

For band gaps, the MLP reach high fidelity with $R^2=0.956$ and sub-0.1~eV median absolute errors (MAE~$=0.087$~eV; Table~\ref{tab:mlp_metrics}). Parity scatter in Figure~\ref{fig:mlp-parity-all} is tightly clustered along the diagonal, and the modest train-test gap ($R^2_{\rm train}=0.982$ vs.\ $R^2_{\rm test}=0.956$) indicates controlled variance and good generalisation. For $S_n$, the model preserves strong rank order ($R^2{\rm test}=0.895$ reported in Table~\ref{tab:mlp_metrics}) with RMSE values of $73.6$. The slightly broader residual spread visible in Figure~\ref{fig:mlp-parity-all} for small $|S|$ is consistent with the known heterogeneity and skew/bimodality of Seebeck distributions in chemically diverse sets.

The band gap accuracy compares favourably with classic tree ensembles and earlier deep models: our MAE and RMSE are on par with, or better than, LightGBM/Random Forest baselines on related datasets and markedly better than earlier MLPs on 2D sets (RMSE $\sim$0.47~eV); see Table~\ref{tab:literature_comparison} for context.\cite{Djeradi2024,Zhang2021} For Seebeck coefficients, our absolute errors (MAE $\approx 39$--$46~\mu$V/K; Table~\ref{tab:mlp_metrics}) are slightly higher than the best-in-class boosted ensembles and specialised deep nets reporting $\sim$20--$37~\mu$V/K.\cite{Furmanchuk2018,BenKamri2024,Antunes2023,Yuan2022} However, those studies often target narrower chemistries or employ tailored feature engineering, whereas our single MLP--trained on a broad, mixed set using a unified representation--reaches consistent performance across both band gap and $S_n$ with limited tuning (Table~\ref{tab:mlp_metrics}, Figure~\ref{fig:mlp-parity-all}). This makes it a robust and reproducible baseline for subsequent interpretability--focused models.

The architecture used here offers a compact parameterisation on the order of $\approx 2.4\times 10^4$ parameters (Table~\ref{tab:params_comparison}) that captures the relatively smooth structure--property relation for band gaps, while being slightly less suited describe nonlinearities that govern the Seebeck  response.  The small train--test deltas across all targets (Table~\ref{tab:mlp_metrics}) suggest neither severe underfitting nor memorisation; instead, the residual errors for $S_n$ likely reflect physics not directly encoded in the features (e.g., carrier concentration), rather than deficiencies in optimisation. 

As a baseline, the MLP coupled to CrystalFormer embeddings delivers (i) state-of-the-art band-gap accuracy relative to general-purpose baselines and (ii) competitive, stable Seebeck predictions on a chemically diverse dataset. This establishes a reliable reference for the KAN models assessed in the next section.

\begin{table}[htbp]
\centering
\caption{Performance metrics (train and test sets) for MLP models trained with CrystalFormer features on the filtered 15,000-sample dataset.}
\label{tab:mlp_metrics}
\begin{tabular}{llcccc}
\hline
\textbf{Property} & \textbf{Set} & \textbf{R$^2$} & \textbf{MSE} & \textbf{RMSE} & \textbf{MAE} \\
\hline
Band Gap   & Train & 0.982 & 0.0113  & 0.106  & 0.065  \\
Band Gap   & Test  & 0.956 & 0.0225  & 0.150  & 0.087  \\
$S_n$      & Train & 0.909 & 4512.4  & 67.17  & 34.44  \\
$S_n$      & Test  & 0.895 & 5473.7  & 73.62  & 38.53  \\
\hline
\end{tabular}
\end{table}

\begin{table}[ht]
\centering
\caption{Parameter breakdown for a conventional MLP and a KAN. 
For the MLP, parameters are separated into edge weights and biases per layer. 
For the KAN, the decomposition shows the actual trainable parameter classes reported by \texttt{PyKAN}.}
\setlength{\tabcolsep}{5pt}
\renewcommand{\arraystretch}{1.05}
\begin{tabularx}{\columnwidth}{llr}
\hline
\textbf{Model} & \textbf{Component} & \textbf{Count} \\
\hline
\multicolumn{3}{l}{\textbf{MLP (128--128--64--1)}} \\
 & Edges (In$\to$128)   & 16,384 \\
 & Bias  (128)          & 128 \\
 & Edges (128$\to$64)   & 8,192 \\
 & Bias  (64)           & 64 \\
 & Edges (64$\to$1)     & 64 \\
 & Bias  (1)            & 1 \\
 & \textbf{Total}       & \textbf{24,833} \\
\hline
\multicolumn{3}{l}{\textbf{KAN (128--16--1)}} \\
 & Spline coefficients  & 30,960 \\
 & Grid/knots           & 2,736 \\
 & Affine parameters    & 12,418 \\
 & Bias                 & 34 \\
 & Other                & 4,128 \\
 & \textbf{Total}       & \textbf{50,276} \\
\hline
\end{tabularx}
\label{tab:params_comparison}
\end{table}

\subsection{Kolmogorov--Arnold Networks for Descriptor Discovery and Reverse Engineering}

After performing a grid search with CrystalFormer descriptors (using $R^2$ and MSE as reference metrics), we selected KAN architectures of $[128,16,1]$ across the three targets.
As summarised in Table~\ref{tab:params_comparison}, our KANs involve more parameters per connection, though by reducing the number of hidden units and layers their overall size can be kept comparable to that of MLPs.  

Despite this comparable parameter count, training KANs is markedly slower. 
In MLPs, forward and backward propagation reduce to highly optimised matrix multiplications and outer products, operations that can scale efficiently on GPUs. 
KANs, however, demand the evaluation of spline basis functions and their derivatives for each edge. 
With $G{+}k=15$ coefficients per connection ($G=12$ grid points, $k=3$ spline order), every pass requires costly polynomial interpolation and gradient calculations. 
Furthermore, continuity constraints couple neighbouring spline coefficients, complicating optimisation and increasing memory overhead by storing spline activations and their derivatives. 
Together, these factors explain why KAN training is computationally more demanding, even when the number of parameters is comparable to MLPs.  

On a multi-core CPU with optimised libraries for dense linear algebra, we observed  significant differences: a baseline MLP with $\sim$24k parameters converged within minutes, whereas a KAN of comparable size ($\sim$50k parameters, cubic splines with $G{+}k=15$) required on the order of eight hours. In effect, the per-parameter cost of KAN training was over two orders of magnitude higher, reflecting the computational burden of spline evaluations and their coupled optimisation. We alleviated this overhead by pruning redundant connections and adopting smaller initial grids, which reduced training time without compromising predictive accuracy.

\begin{figure}[htbp]
  \centering

  \includegraphics[width=0.5\textwidth]{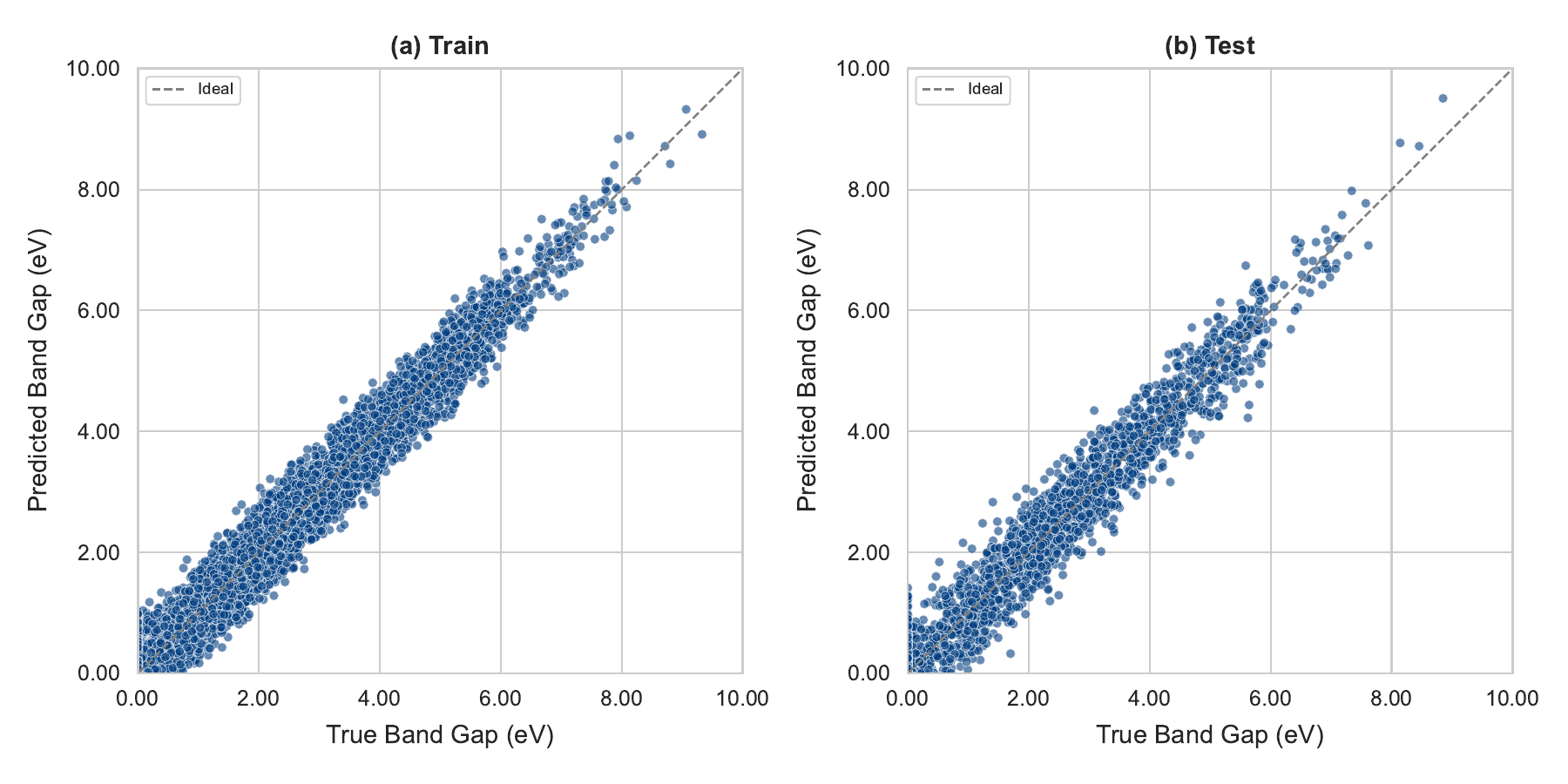} \\
  \includegraphics[width=0.5\textwidth]{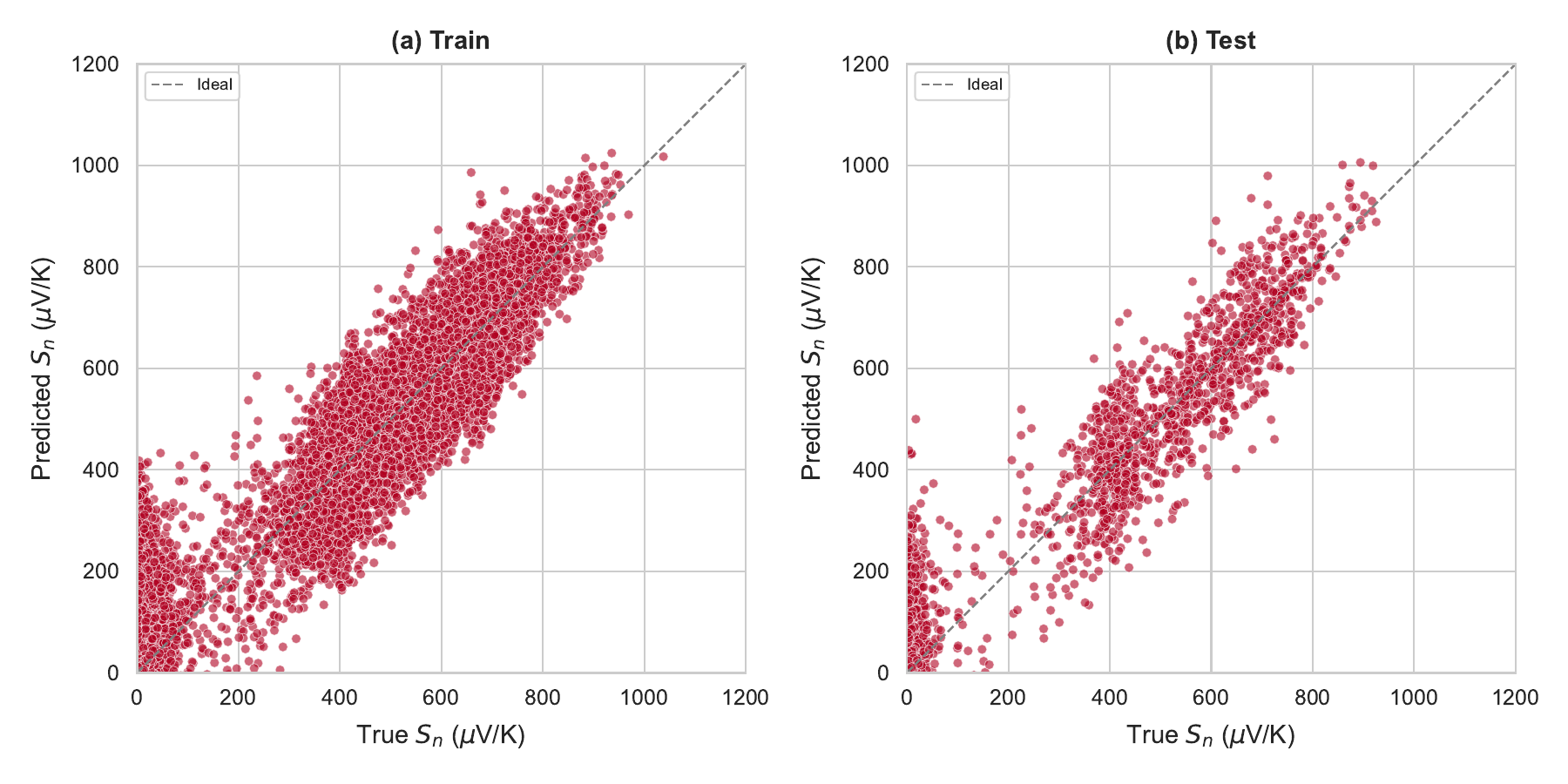} \\
  \caption{Parity plots for predicted vs. true values of the band gap and $S_n$ (top and bottom, respectively) using the KAN model. Each property is shown for both training and test sets (left to right). The diagonal line represents perfect prediction.}
  \label{fig:kan_parity}
\end{figure}

The trained KAN models achieved consistently high predictive performance across band gap and Seebeck coefficients, with parity plots (Fig.~\ref{fig:kan_parity}) showing strong agreement between predicted and true values in both training and test sets.
The performance metrics are summarised in Table~\ref{tab:kan_metrics}, demonstrating that KANs performance is comparable to  multilayer perceptrons (Table~\ref{tab:mlp_metrics}), albeit with the added benefit of interpretability. 
In particular, the accuracy levels obtained for band gap exceed MLP whereas the Seebeck predictions are slightly below the best-performing MLP baselines, yet remain competitive with recent state-of-the-art machine learning approaches reported in the literature (see Table~\ref{tab:literature_comparison}), including gradient-boosted decision trees, deep neural networks, and graph-based models. 
These results establish KANs as a viable alternative to conventional architectures.

\begin{table}[htbp]
\centering
\caption{Performance metrics (train and test sets) for KAN models trained with CrystalFormer features on the filtered 15,000-sample dataset.}
\label{tab:kan_metrics}
\begin{tabular}{llcccc}
\hline
\textbf{Property} & \textbf{Set} & \textbf{R$^2$} & \textbf{MSE} & \textbf{RMSE} & \textbf{MAE} \\
\hline
Band Gap   & Train & 0.974 & 0.0225  & 0.112  & 0.072  \\
Band Gap   & Test  & 0.968 & 0.0324  & 0.146  & 0.094  \\
$S_n$      & Train & 0.895 & 5260.7  & 71.27  & 35.12  \\
$S_n$      & Test  & 0.851 & 7085.1  & 69.32  & 39.09  \\
\hline
\end{tabular}
\end{table}

\begin{table*}[htbp]
\centering
\caption{Representative machine learning models reported between 2015 and 2025 for predicting band gaps and Seebeck coefficients of inorganic materials. Results include both recent models (2023--2025) and selected high-performing baselines from earlier studies.}
\label{tab:literature_comparison}

\setlength{\tabcolsep}{4pt}
\renewcommand{\arraystretch}{1.05}

\begin{tabular*}{\textwidth}{@{\extracolsep{\fill}}llcccc}
\hline
\textbf{Target property} & \textbf{Model type} & \textbf{$R^2$} & \textbf{MSE} & \textbf{RMSE} & \textbf{MAE} \\
\hline
Band gap (2D materials)               & GBDT \cite{Zhang2021}                         & 0.92  & --            & 0.24 eV     & --           \\
Band gap (2D materials)               & MLP \cite{Zhang2021}                          & 0.70  & --            & 0.47 eV     & --           \\
Band gap (perovskites)                & LightGBM \cite{Djeradi2024}                   & 0.934 & --            & --          & 0.302 eV     \\
Band gap (perovskites)                & XGBoost \cite{Djeradi2024}                    & 0.911 & --            & --          & 0.350 eV     \\
Band gap (perovskites)                & Random Forest \cite{Djeradi2024}              & 0.921 & --            & --          & 0.320 eV     \\
Band gap (perovskite oxides)          & Ensemble model \cite{Talapatra2023}          & 0.86  & $\sim$0.07 eV$^2$ & $\sim$0.26 eV & 0.18 eV \\
Band gap                              & SVR/GBDT + SISSO \cite{Huo2024}               & --    & --            & 0.36 eV     & --           \\
Band gap                              & CGCNN (domain adaptation) \cite{Haghshenas2024} & -- & --            & --          & 0.23 eV      \\
Band gap (mixed materials)            & CrystalFormer \cite{Wei2024}                  & 0.97  & --            & 0.048 eV    & 0.033 eV     \\
Band gap (semiconductors)             & GNN + spectral features \cite{Cheng2024}      & 0.945 & --            & 0.11 eV     & 0.08 eV      \\
Band gap (inorganics)                 & Deep KRR + SOAP \cite{Xie2023}                & 0.89  & --            & --          & 0.22 eV      \\
\hline
Seebeck ($S_n$/$S_p$)                 & CraTENet \cite{Antunes2023}                   & 0.78  & --            & --          & $\sim$114 $\mu$V/K \\
Seebeck ($S_n$/$S_p$)                 & RF \cite{Antunes2023}                         & 0.79  & --            & --          & $\sim$141 $\mu$V/K \\
Seebeck ($S_n$/$S_p$)                 & CraTENet+gap \cite{Antunes2023}               & 0.96  & --            & --          & $\sim$49 $\mu$V/K  \\
Seebeck ($S_n$/$S_p$)                 & RF+gap \cite{Antunes2023}                     & 0.96  & --            & --          & $\sim$54 $\mu$V/K  \\
Seebeck ($S_n$/$S_p$)                 & NN + elemental features \cite{Yuan2022}       & 0.96  & --            & --          & 31--39 $\mu$V/K    \\
Seebeck ($S_p$)                       & GBT/CatBoost \cite{Furmanchuk2018}            & 0.73  & --            & 85          & 55 $\mu$V/K       \\
Seebeck ($S_p$, half-Heusler)         & XGBoost ensemble \cite{BenKamri2024}          & 0.95  & --            & --          & 20.8 $\mu$V/K     \\
Seebeck ($S_n$, half-Heusler)         & LightGBM ensemble \cite{BenKamri2024}         & 0.94  & --            & --          & 20.8--37.0 $\mu$V/K \\
Seebeck (mixed, exp.\ data)$^{\dagger}$ & XGBoost \cite{Na2022}                        & 0.90  & --            & --          & 21.1 $\mu$V/K     \\
Seebeck (inorganics)                  & Matminer baseline \cite{don2024predictive}    & 0.85  & --            & --          & 36.70 $\mu$V/K    \\
Seebeck (inorganics)                  & GNN (symmetry-aware) \cite{Cheng2024}         & 0.93  & --            & --          & 19--21.7 $\mu$V/K \\
\hline
\end{tabular*}

\begin{flushleft}
{\small $^{\dagger}$ Trained on a broad experimental thermoelectric dataset of 5{,}205 samples; model achieved $R^2 \geq 0.90$ for multiple transport properties.\\
\textbf{Note:} ``--'' indicates that the metric was not explicitly reported. Errors are in eV for band gap and in $\mu$V/K for the Seebeck coefficient.}
\end{flushleft}

\end{table*}

\subsection{Robustness and generalization across chemical space}

\begin{figure}[t]
  \centering
  \includegraphics[width=0.48\textwidth]{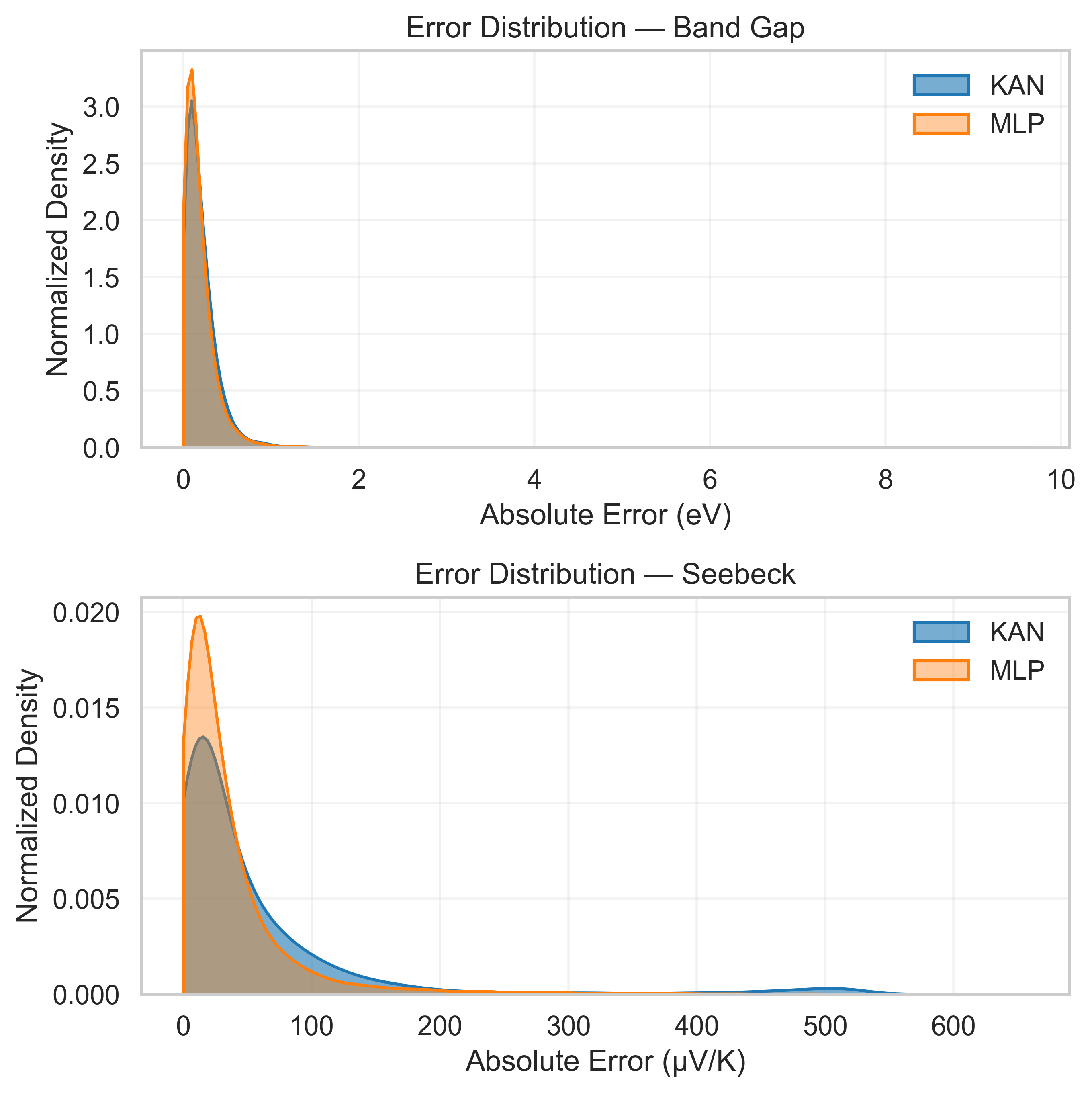}
  \caption{
  Kernel density estimates (KDE) of absolute prediction errors for (top) electronic band gap and (bottom) Seebeck coefficient, comparing Kolmogorov--Arnold Networks (KAN, blue) and Multilayer Perceptrons (MLP, orange).}
  \label{fig:error_dists}
\end{figure}

The predictive behaviour of the KAN and MLP architectures was evaluated across the entire dataset using both global and element-resolved diagnostics. 
Figure~\ref{fig:error_dists} compares the absolute error distributions for electronic band gap and Seebeck coefficient. 
In both properties, the error density is sharply centred at low values, confirming that both models achieve consistent regression behaviour across most compounds. 
The KAN exhibits a slightly lower central peak and a small secondary peak at large error values compared with the MLP, indicating a broader but more stable distribution.

\begin{figure*}[t]
    \centering
    \includegraphics[width=\textwidth]{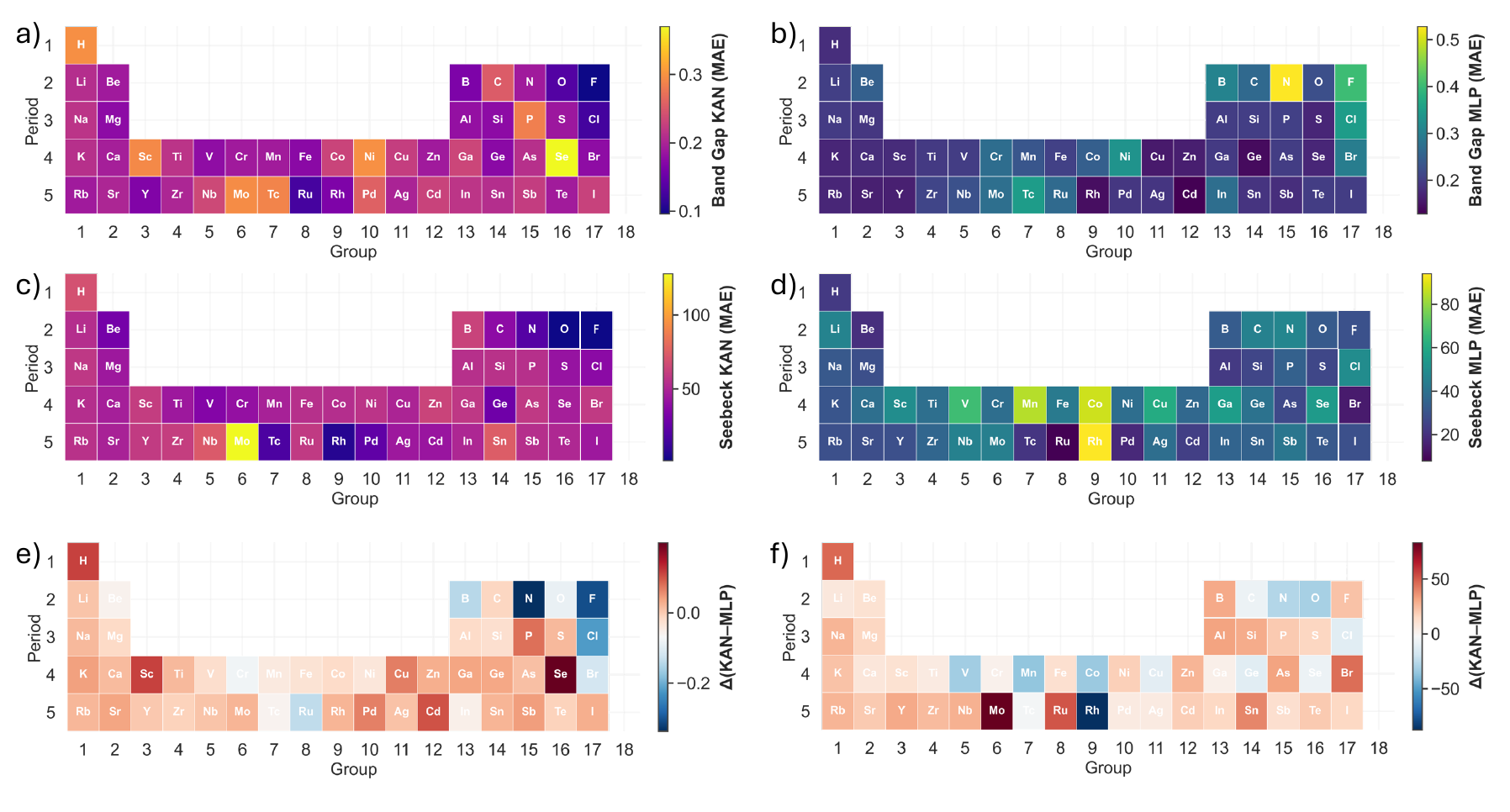}

\caption{
\textbf{Element-resolved mean absolute error (MAE) maps for band gap and Seebeck predictions.}
Each periodic-table cell represents the mean error across all compounds containing that element.
Panels (a-b) show element-wise MAE for the KAN and MLP band-gap predictions, while (c-d) report analogous values for Seebeck coefficients.
Panels (e-f) display the inter-model differences $\Delta$(KAN-MLP), where blue tones indicate superior KAN performance.
KAN achieves more uniform accuracy across s-, p-, and d-block elements, while MLP exhibits stronger element-dependent variability, particularly for transition metals and chalcogens.}
\label{fig:periodic_mae}
\end{figure*}

To identify the chemical origins of uncertainty, the mean absolute error (MAE) and inter-model difference 
$\Delta = \mathrm{Value}_\mathrm{KAN} - \mathrm{Value}_\mathrm{MLP}$ were averaged for each element across all compounds 
containing that species (Fig.~\ref{fig:periodic_mae}). 
Here, $\mathrm{Value}_\mathrm{KAN}$ and $\mathrm{Value}_\mathrm{MLP}$ denote the property values predicted by the 
Kolmogorov--Arnold Network and the Multilayer Perceptron, respectively, and $\Delta$ represents their pointwise difference. 
Both properties reveal systematic chemical trends. 

For band gaps, the largest errors are observed in compounds containing Se (KAN) or N (MLP), as well as those including 
late transition metals (Fe, Co, Ni, Cu) and $p$-block elements. On average, the KAN performs best for halogens, confirming 
reliable learning in simple ionic regimes, whereas the MLP performs worst in these same groups. Conversely, in alkali and 
alkaline-earth metals, the MLP achieves slightly lower errors. 

For Seebeck coefficients, the overall trend differs from that of the band gap: the largest errors occur for Mo (KAN) and 
Rh (MLP). In general, the MLP exhibits marginally lower element-wise variance. Consequently, the difference maps 
$\Delta(\text{KAN--MLP})$ are predominantly red (positive), indicating that the MLP generally produces lower element-averaged 
MAEs. However, both models maintain low absolute errors, and the element-wise differences are small relative to the overall 
prediction magnitude, confirming that generalisation is robust for both architectures across most elements. 
This highlights that, for Seebeck prediction, KAN’s superior robustness arises not from mean accuracy, but from its 
ability to prevent physically inconsistent predictions 
(Table~\ref{tab:KAN_MLP_disagreement_structured}).

\subsubsection{Origin of Predictive Uncertainty}

By comparing the elements with the largest and smallest mean absolute errors ($\text{MAE}_{\text{avg}}$ across KAN and MLP) in band-gap prediction, we can trace the physical origin of these differences and clarify how KAN and MLP handle distinct bonding and electronic regimes.

\paragraph{High-Uncertainty Elements (Large $\text{MAE}_{\text{avg}}$):} 
The elements exhibiting the highest prediction errors (e.g., $\text{N, C, F, P, Se}$ in the $p$-block; $\text{Ni, Co, Mo, Tc}$ in the $d$-block) share two defining characteristics:
\begin{enumerate}
    \item \textbf{Covalent Directionality and Localisation:} 
    Light $p$-block elements ($\text{N, C, F}$) form highly directional $\sigma$ and $\pi$ bonds, requiring the model to resolve intricate, non-linear dependencies on bond angles and orbital overlap. This leads to strong localisation of valence states and sensitivity of the band gap ($E_g$) and Seebeck coefficient ($S$) to small structural deviations. KAN shows smoother regression behavior and fewer extreme outliers in these cases, while MLP tends to produce abrupt prediction jumps.
    \item \textbf{Partially Filled $d$-Shells and Correlation:} 
    Mid-to-late transition metals ($\text{Ni, Co, Mo, Tc}$) introduce partially filled $d$-orbitals, which leads to strong $\text{d--d}$ electron correlation effects (Mott-Hubbard physics) or highly complex $\text{p-d}$ hybridization near the Fermi level. These phenomena are intrinsically difficult for standard machine learning models, whose embeddings struggle to fully parameterize the non-local and dynamic electronic interactions. Here, KAN's locally adaptive spline basis mitigates overfitting by maintaining smooth interpolation across correlated regions, whereas MLP exhibits larger error variance and occasional discontinuities.
\end{enumerate}

\paragraph{Low-Uncertainty Elements (Low $\text{MAE}_{\text{avg}}$):} 
Conversely, the elements associated with the lowest errors (e.g., $\text{Mg, Sr, Rb}$ in the $s$-block; $\text{Zn, Cd}$ in the $d$-block; $\text{O, S, Ge}$ in the $p$-block) are characterized by electronic simplicity and predictability:
\begin{enumerate}
    \item \textbf{Simple Ionicity:} Alkali and alkaline-earth metals drive simple ionic bonding in wide-gap insulators. Their contributions to the band structure are often far from the Fermi level, resulting in predictable band gaps dominated by well-behaved $\text{s}$ and $\text{p}$ states.
    \item \textbf{Full or Simple Shells:} Elements like $\text{Zn}$ and $\text{Cd}$ possess a full $d^{10}$ shell, rendering their chemistry and electronics less complex than those of their neighbors with partially filled $d$-bands. Likewise, the common, highly represented structures formed by $\text{O}$ and $\text{S}$ (simple oxides and sulfides) provide abundant and consistent training examples, enabling robust model learning.
\end{enumerate}

This analysis suggests that the primary challenge for the predictive models lies not in simple compositional variety, but in regions of the periodic table defined by high electronic correlation and highly directional, non-linear bonding characteristics. This robustness in modeling highly nonlinear functional dependencies makes KAN uniquely suited for the subsequent step of symbolic regression, where stable and smooth functional forms are crucial for physical interpretability.

\begin{figure*}[t]
\centering

 \includegraphics[width=0.45\linewidth]{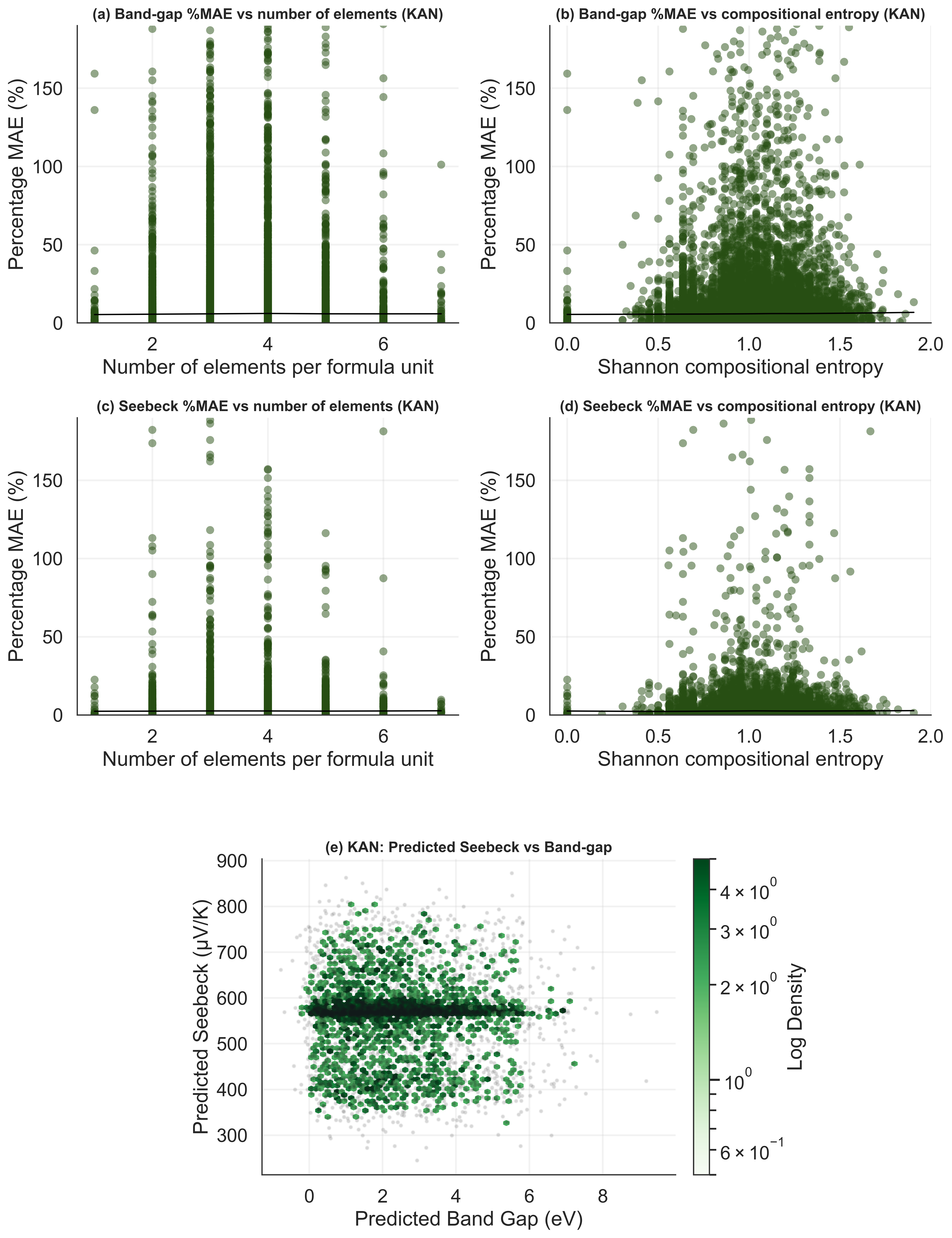} 
 \hfill
  \includegraphics[width=0.45\linewidth]{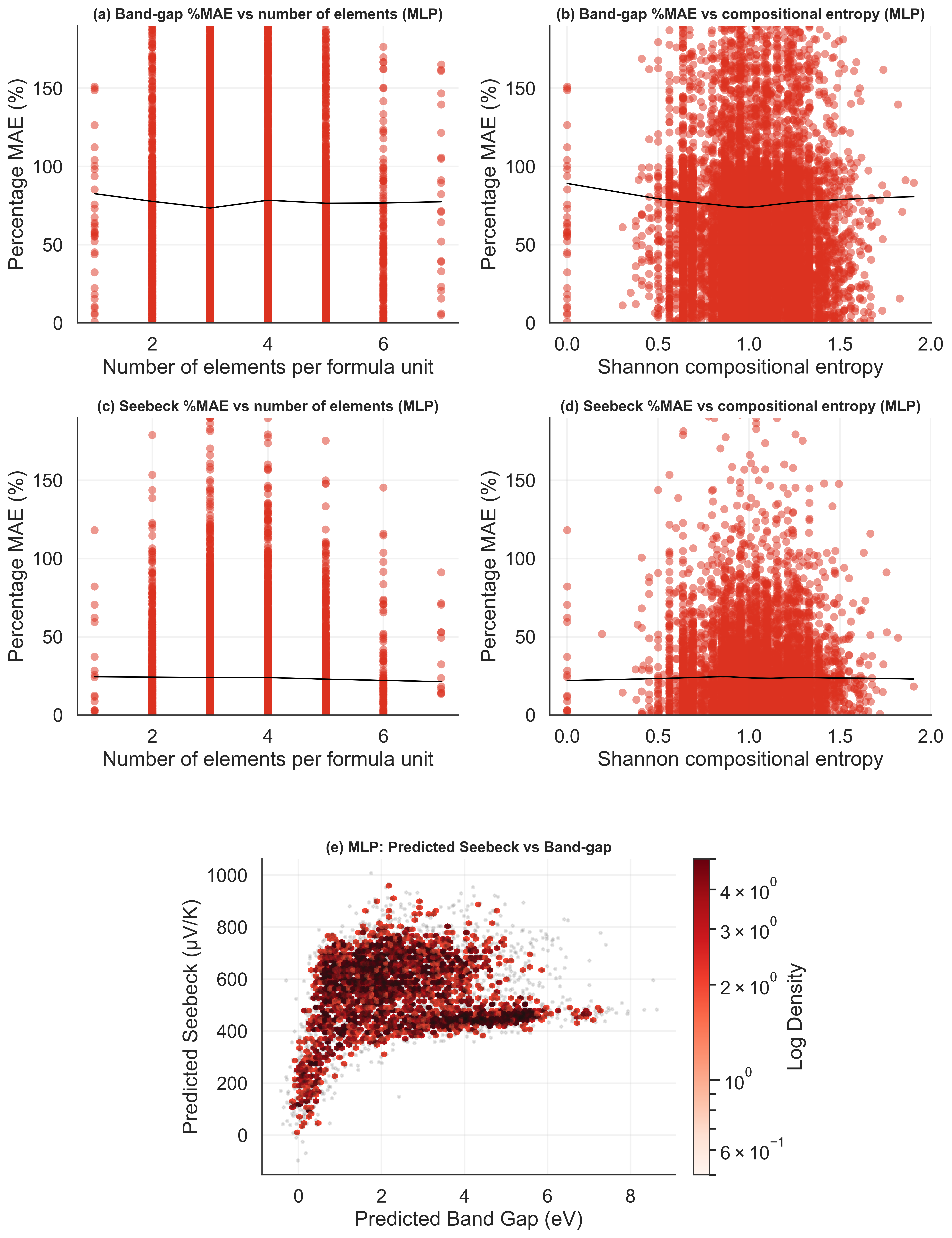}

\caption{
\textbf{Error behaviour across compositional complexity and electronic regimes.} 
Panels left (green) and right (red) compare the performance of KAN and MLP models, respectively, as functions of compositional entropy and elemental diversity, showing that both generalise well across simple and high-entropy materials. 
At high complexity, MLP predictions become more scattered, whereas KAN maintains smoother, composition-invariant accuracy. 
The lower panels illustrate the relationship between Seebeck coefficient and band gap: MLP follows a single dominant Pisarenko-like trend, while KAN displays a broader but more continuous mapping that captures multiple transport regimes within a unified functional representation.
}
\label{fig:complexity_density}
\end{figure*}

To probe whether chemical complexity correlates with predictive uncertainty, we examined the percentage MAE as a function of both the number of elements per formula unit and the Shannon compositional entropy, 
$S_\mathrm{chem} = -\sum_i x_i \ln x_i$, where $x_i$ denotes the fractional atomic concentration (see Fig.~\ref{fig:complexity_density}). 
The parameter $n_\mathrm{el}$ indicates the number of distinct elements per formula unit. 
Spearman's rank correlation coefficient, $\rho$, is used to quantify monotonic relationships between model errors and compositional descriptors.
No systematic dependence was observed ($|\rho| \leq 0.05$ across all properties), confirming that both architectures generalise 
robustly across binary, ternary, and high-entropy materials without accuracy degradation, with a more pronounced trend of KAN 
performing slightly better at the extremes of compositional complexity. 

The MLP shows broader dispersion across entropy and elemental diversity, whereas KAN maintains narrower variance. 
This behaviour reflects KAN’s functional regularisation, which constrains local oscillations and preserves compositional 
smoothness under sparse sampling. 

When the Seebeck coefficient is plotted against the band gap (Fig.~\ref{fig:complexity_density}, bottom panels), both models 
reproduce the general Pisarenko-like trend where $S$ increases with $E_g$, but the KAN exhibits a weaker global correlation 
(Spearman $\rho = -0.02$) compared to the MLP ($\rho = 0.11$). 
The reduced conditional spread of Seebeck values at fixed band gap confirms that KAN produces smoother and more globally 
consistent structure-property mappings. However, the flatter overall $E_g$-$S$ trend also suggests that spline 
regularisation, while stabilising predictions, may smooth out subtle but physically meaningful nonlinear correlations -- 
particularly in sparsely sampled or mixed-conduction regimes.

A more targeted comparison was performed by analysing the 10\% of compounds 
with the largest KAN--MLP prediction differences. 
Statistical analysis (Partially reported in Tab.~\ref{tab:KAN_MLP_disagreement_structured}) shows that 
high-disagreement band-gap compounds possess slightly larger average gaps but 
similar compositional complexity ($p_{S_\mathrm{chem}}=0.043$, $p_{n_\mathrm{el}}=0.26$). 
Here, $p_{S_\mathrm{chem}}$ and $p_{n_\mathrm{el}}$ denote the $p$-values obtained 
from two-sample tests comparing the Shannon entropy and element-count distributions between the full and high-disagreement subsets.

For Seebeck coefficients, both the entropy and element-count distributions 
are statistically indistinguishable ($p \geq 0.1$),  suggesting that the 
disagreement may arise from intrinsic electronic factors rather than structural 
diversity, consistent with the band-gap findings.
Importantly, in the high disagreement region, KAN performs consistently better than MLP with respect to the true target values.

The convergence of these analyses establishes three key findings:
\begin{enumerate}
    \item The largest inter-model divergences occur in wide-bandgap systems with 
    strong orbital localisation and low carrier density near the band edges.
    \item KAN exhibits smoother and more stable mappings in sparsely sampled or 
    highly nonlinear regimes, while MLP predictions can be discontinuos.
    \item For thermoelectric properties, both models generalise well across diverse  chemistries.
    \item   KAN more effectively suppresses high-error outliers whereas MLP  preserves physically consistent Seebeck--band-gap relationships.
\end{enumerate}
Overall, the results demonstrate that KAN's continuous functional representation 
yields greater robustness and compositional invariance, especially in complex materials where smooth structure--property relationships are essential for reliable screening.

\begin{figure*}[t]
\centering
\includegraphics[width=0.45\textwidth]{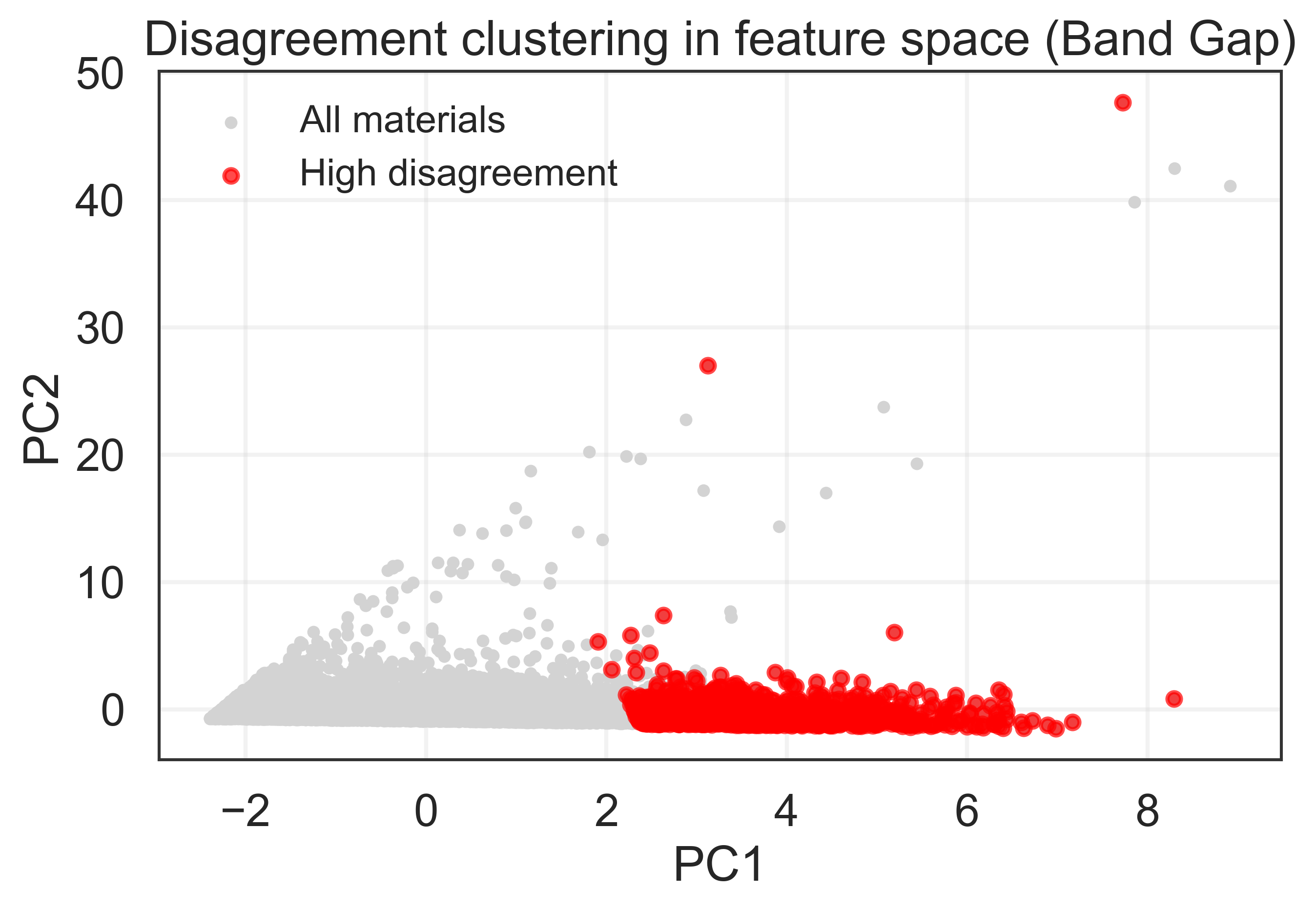}
\includegraphics[width=0.45\textwidth]{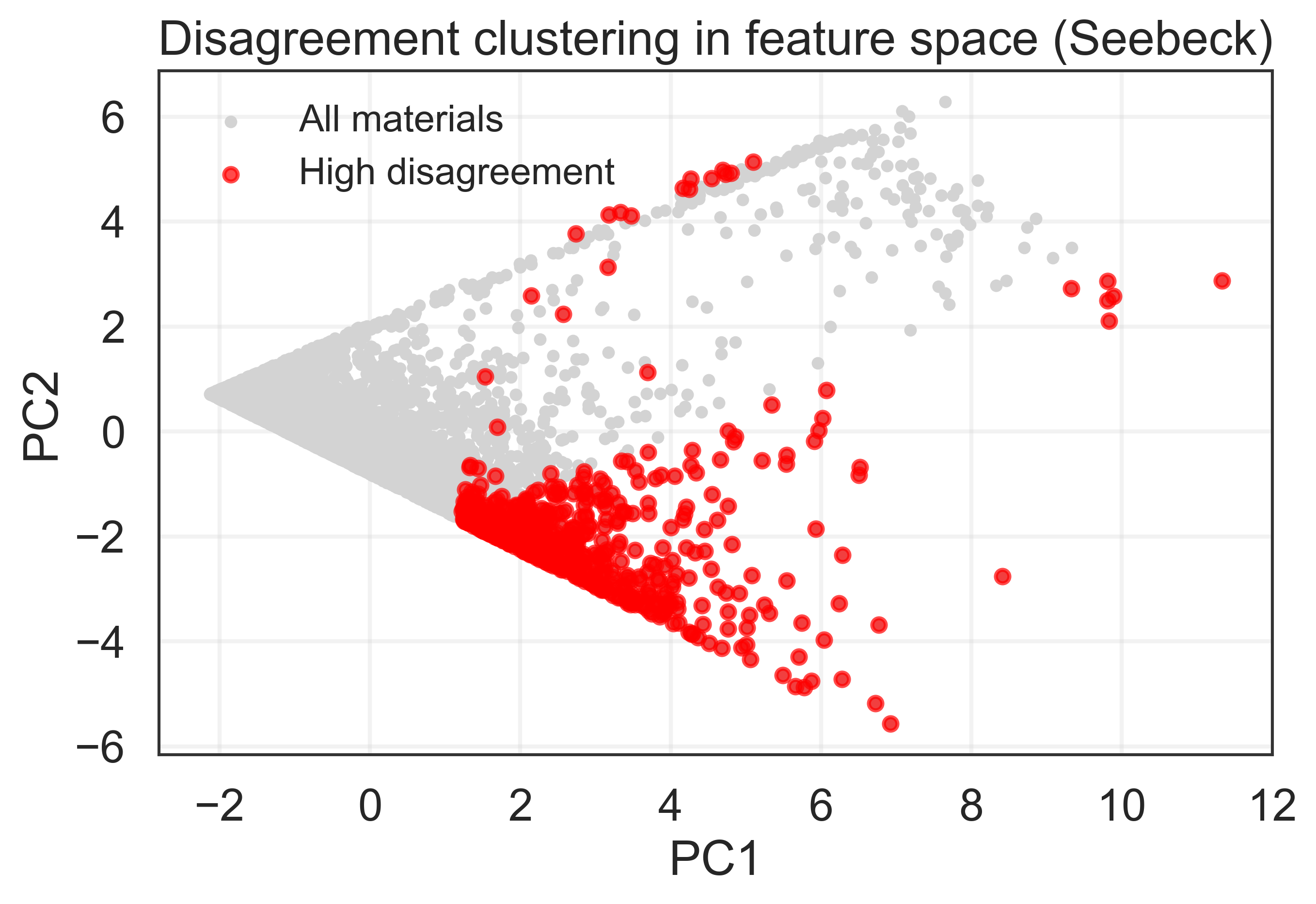}
\caption{
Principal component projections of the Crystalformer embeddings for (left) band-gap and (right) Seebeck datasets.
Red points highlight the 0.2\% tail of highest KAN--MLP disagreement.
}
\label{fig:pca_disagreement}
\end{figure*}

To identify specific regions of feature space associated with large inter-model discrepancies, 
the Crystalformer embeddings were projected onto their first two principal components (PC1 and PC2; Fig.~\ref{fig:pca_disagreement}). 
This two-dimensional feature map shows the entire materials dataset as grey points and the compounds with the highest 
KAN--MLP disagreement as red points. Distinct clusters emerge in chemically complex regions of descriptor space, whereas 
the main data density remains concentrated near the origin ($\text{PC}1 \approx 0, \text{PC}2 \approx 0$), corresponding 
to the chemically simplest and most abundant training instances.

The compounds where KAN and MLP exhibit the largest disagreement in band-gap prediction form a compact, displaced cluster 
($\text{PC}1 \geq 2$), suggesting that failure modes are not random but correlated with regions of higher structural and 
chemical complexity in the Crystalformer feature space. For Seebeck prediction, the disagreement clusters are more dispersed, 
extending across $\text{PC}2$, which indicates that the thermoelectric response is more sensitive to subtle variations in 
chemical environment and bonding topology.

The disagreement clustering for Seebeck prediction shows a greater degree of dispersion than the band-gap failure modes  (Fig.~\ref{fig:pca_disagreement}). While still displaced along $\text{PC}1$ ($\text{PC}1 \geq 1.5$), the Seebeck failure modes spread considerably across the $\text{PC}2$ axis, forming a bifurcated structure extending into both positive and negative $\text{PC}2$ space. This increased $\text{PC}2$ variance indicates that the mechanisms governing $S$ prediction are more acutely sensitive to chemical/structural variations captured by $\text{PC}2$ than the mechanisms governing band-gap prediction. This property-specific dispersion suggest that thermoelectric response is governed by regime-dependent transport behaviors that become numerically separable in the high-complexity regions of chemical space.

\subsubsection{Interpretability and physical consistency.} 
The improved stability of KAN across chemically and electronically complex regimes arises from its architectural inductive bias: a locally adaptive spline basis that enforces continuity in both value and derivative space. 
This smooth functional mapping mitigates discontinuities that often emerge in parametric networks when extrapolating beyond dense training regions, ensuring physically plausible interpolation between sparse data points. 
In practice, this manifests as fewer abrupt shifts in predicted band structures and transport coefficients, especially in wide-gap and strongly localised systems. 
The resulting continuity not only enhances predictive robustness but also facilitates interpretability.
The learned spline functions can be directly inspected to reveal local descriptor--property dependencies -- a feature absent in conventional black-box MLPs.

Notably, KAN produces no sign inversions and recovers correct electronic regimes in all cases, confirming that its spline-based continuity prevents the erratic extrapolations characteristic of MLP failures. 
These compounds serve as diagnostic markers: they pinpoint precisely where KAN's locally adaptive, continuous representation enforces physical plausibility while MLP's discrete function approximation breaks down. 
All materials shown are from the upper 0.2\% tail of the error distribution, representing the most challenging test cases for generalization.  

Across all the worst-performing compounds, the KAN consistently delivers smaller absolute errors than the MLP, confirming that its spline-based formulation not only improves average stability but also dominates in the most challenging prediction regimes.  

A concise summary of these critical cases is provided in Table~\ref{tab:KAN_MLP_disagreement_structured}.


\begin{table*}[t]
  \centering
  \caption{
  \textbf{Representative materials exhibiting the largest KAN--MLP prediction discrepancies.}
  Each compound is listed with the model providing the most accurate result and a brief description of its key physical or chemical characteristics. 
  These cases illustrate the diversity of bonding types and electronic regimes represented in the dataset.
  }
  \vspace{4pt}
  \scriptsize
  \renewcommand{\arraystretch}{0.92}
  \setlength{\tabcolsep}{2pt}
  \begin{ruledtabular}
  \begin{tabular}{lcl}
  \multicolumn{3}{c}{\textit{Band-gap predictions}} \\
  \hline
  \textbf{Compound} & \textbf{Best model} & \textbf{Main material characteristics} \\
  \hline
  TmMg$_5$                   & KAN & Intermetallic compound with partially filled $f$-shell (Tm) and metallic bonding; narrow-gap or semimetallic behaviour. \\
  CaIn$_2$                   & KAN & Polar semimetal with covalent In--In chains and metallic Ca layers; features delocalised $p$-states. \\
  Ga$_3$Co                   & KAN & Metallic compound with strong $p$--$d$ hybridisation between Ga and Co; representative of covalent--metallic alloys. \\
  SrCaI$_4$                  & KAN & Highly ionic mixed halide containing alkaline-earth metals; wide-gap insulator dominated by ionic bonding. \\
  Er$_2$Si$_3$Rh             & KAN & Complex intermetallic silicide with mixed $d$--$f$ electron character and heavy-atom correlation. \\
  B$_2$Mo                    & KAN & Covalent boride with strong B--B $\pi$ bonding and Mo $d$-state hybridisation; narrow-gap semimetal. \\
  SbH(OF$_3$)$_2$            & KAN & Molecular-like oxyfluoride with highly electronegative ligands; polar covalent bonding and strong anion asymmetry. \\
  BaHfMo                     & KAN & Double perovskite-type oxide with mixed $d$-electron occupation (Hf/Mo); moderate band gap and ionic--covalent bonding. \\
  Li$_4$Fe$_3$Co$_3$(WO$_8$)$_2$ & KAN & Multimetallic oxide with spin-polarised Fe/Co $d$-states and strong electron correlation typical of transition-metal oxides. \\
  \hline
  \hline
  \multicolumn{3}{c}{\textit{Seebeck predictions}} \\
  \hline
  \textbf{Compound} & \textbf{Best model} & \textbf{Main material characteristics} \\
  \hline
  CSe$_2$S$_2$N$_2$(OF)$_3$  & KAN & strong covalency and high electronegativity contrast; prone to large carrier localisation. \\
  AuSe                       & KAN & Narrow-gap semimetal with strong spin--orbit coupling and hybridised $d$--$p$ orbitals. \\
  Na$_5$Fe$_2$P$_2$(CO$_7$)$_2$ & KAN & Polyanionic phosphate--carbonate complex containing Fe $d$-states; polaronic conduction and mid-gap thermopower. \\
  Sm$_3$S$_4$                & KAN & Narrow-gap rare-earth sulfide with strongly correlated $f$-electrons; known for high thermoelectric response. \\
  HgTe                       & KAN & Small-gap semimetal and topological material; strong $p$--$d$ mixing and inverted band structure. \\
  SmAs                       & KAN & Rare-earth pnictide with mixed covalent--metallic bonding; $f$--$p$ hybridisation near the Fermi level. \\
  Te$_4$MoBr                 & KAN & Layered telluride--bromide system with mixed ionic and covalent bonding; heavy-atom conduction bands. \\
  Ti$_3$Cu$_2$Te(PO$_4$)$_6$ & KAN & Multimetallic chalcogenophosphate with correlated $d$-electron transport and narrow conduction band. \\
  Ba$_5$Re$_3$ClO$_{15}$     & KAN & Rhenium-based perovskite oxide with extended $d$-band manifold and polar halide coordination. \\
  \end{tabular}
  \end{ruledtabular}
\end{table*}

\subsection{Symbolic Representation}

\begin{figure}[htbp]
    \centering
        \includegraphics[width=0.4\textwidth]{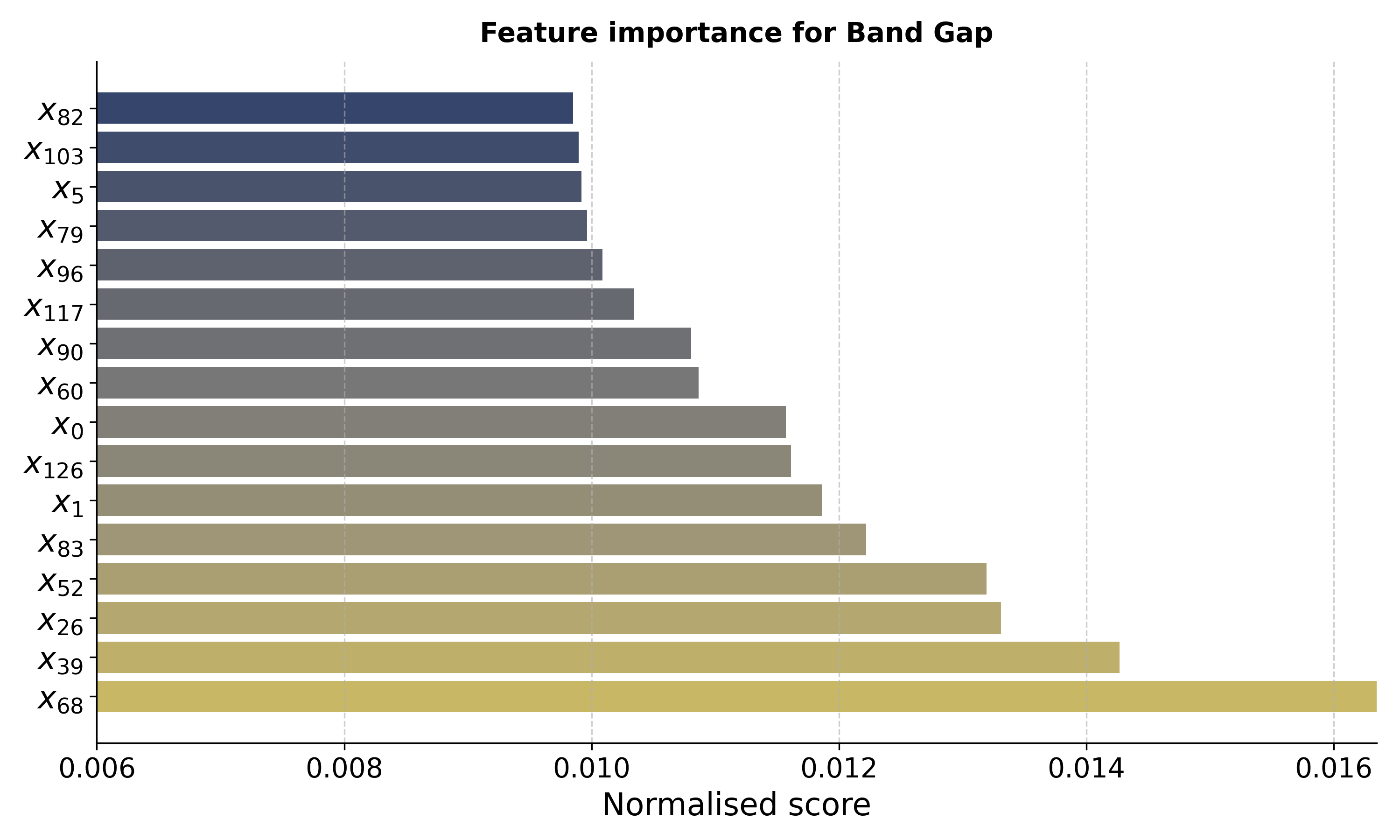} \\
    \includegraphics[width=0.4\textwidth]{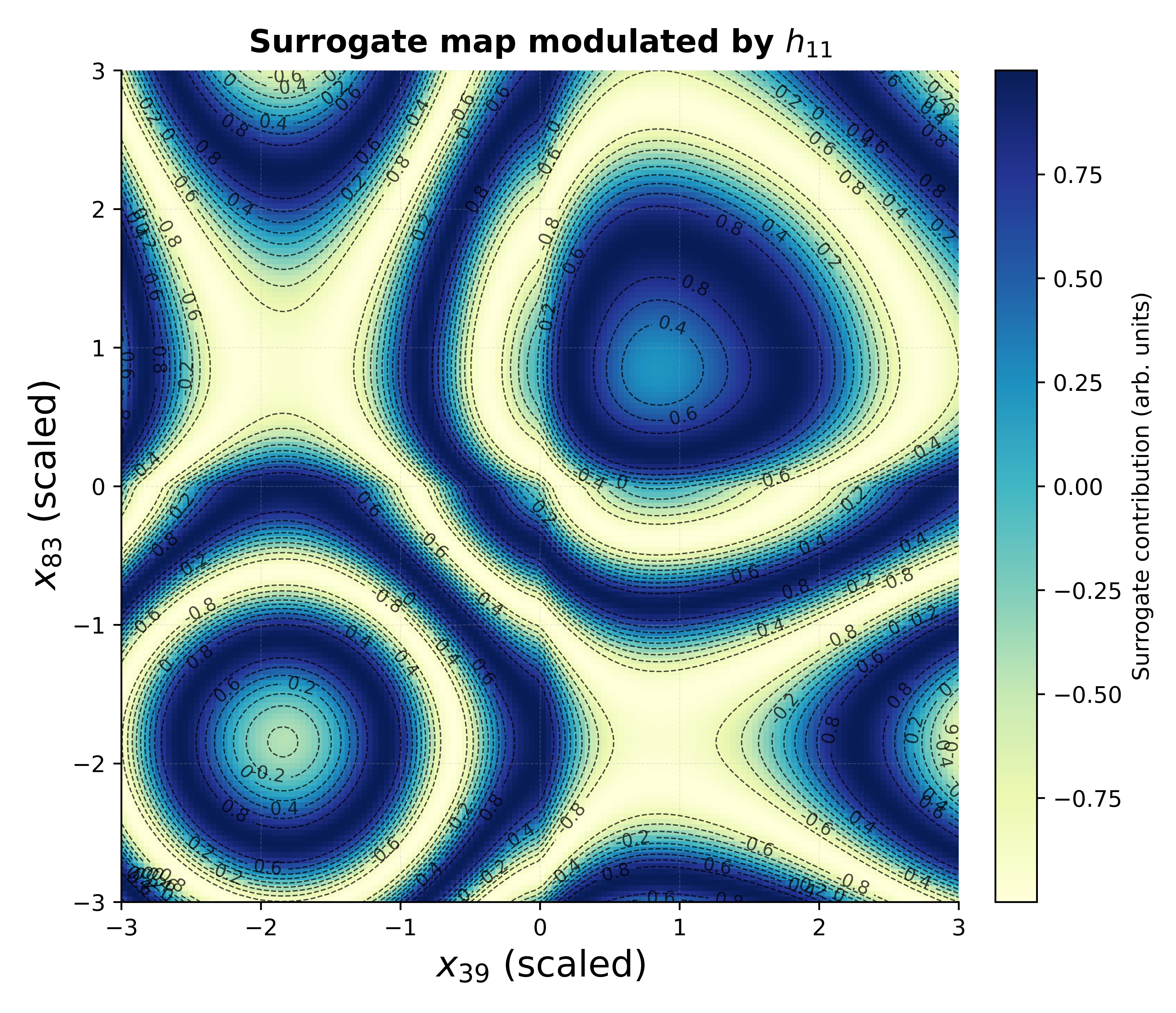} \\
        \includegraphics[width=0.4\textwidth]{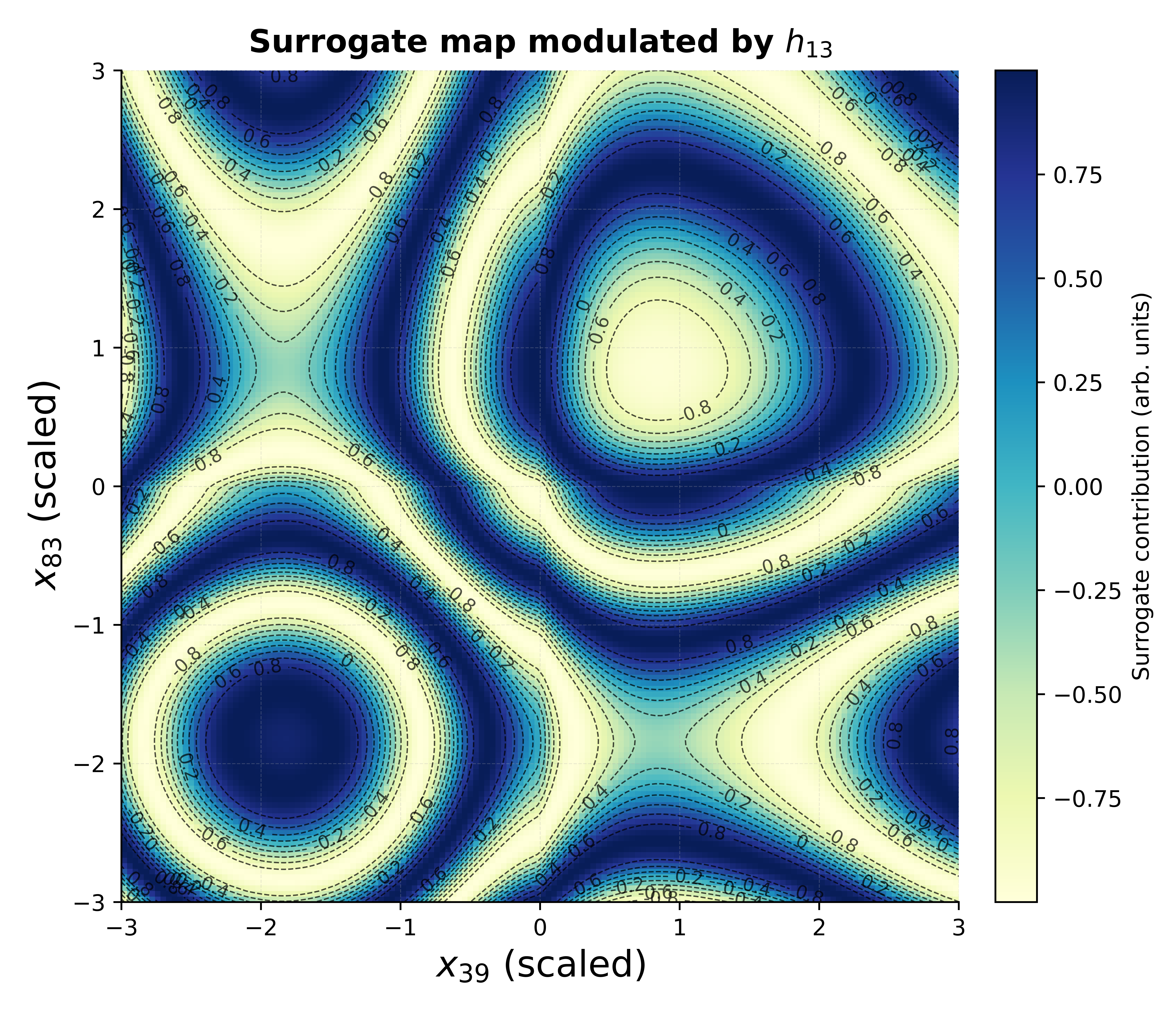} \\
\caption{Top: band gap feature attribution scores computed using the KAN model. The ten descriptors with the highest attribution scores were selected for interpretability analysis. Middle and bottom: Fully modulated surrogate functions for the band gap model, constructed using the most relevant input descriptors.}
    \label{fig:band_gap_attr_contours}
\end{figure}

\begin{figure}[htbp]
    \centering
        \includegraphics[width=0.4\textwidth]{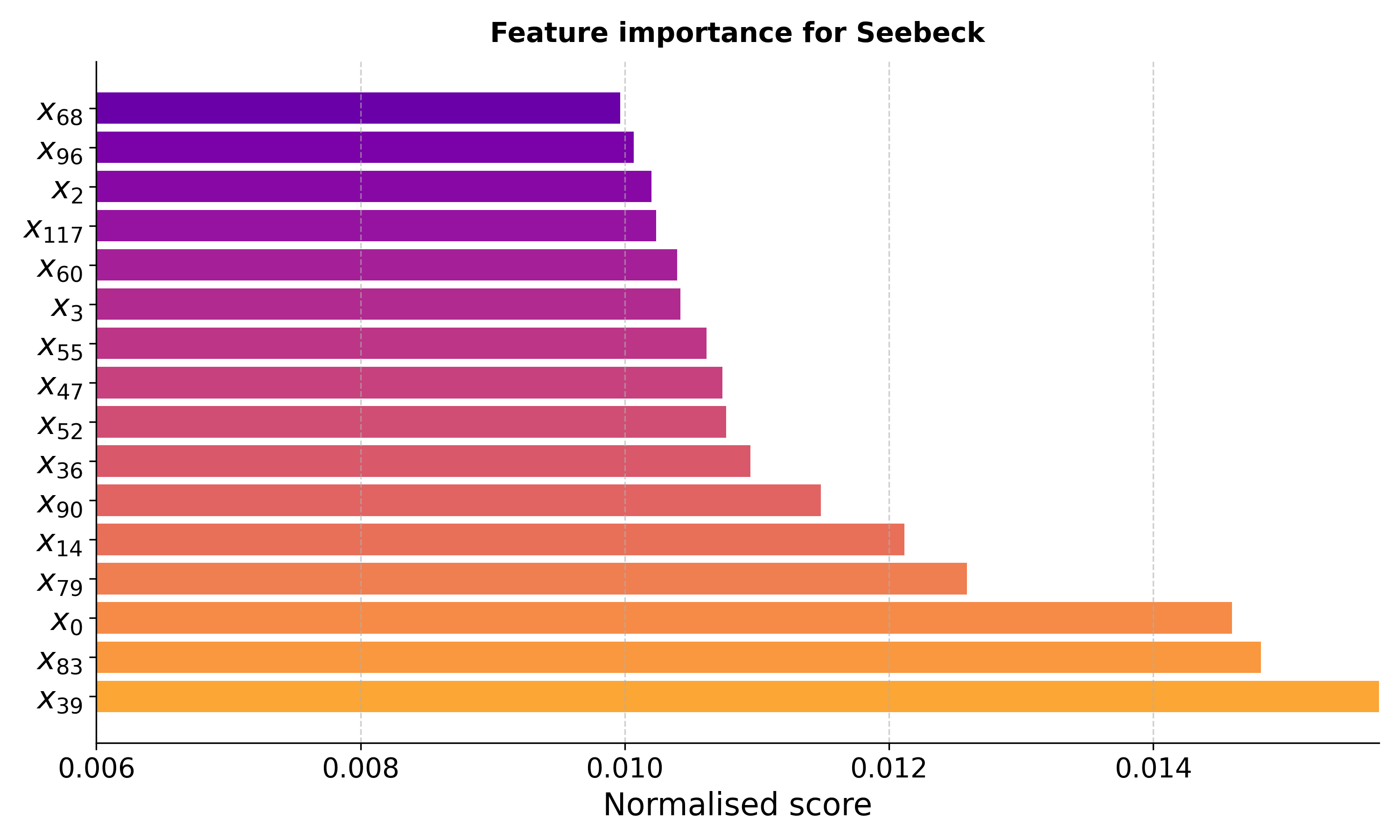} \\
    \includegraphics[width=0.4\textwidth]{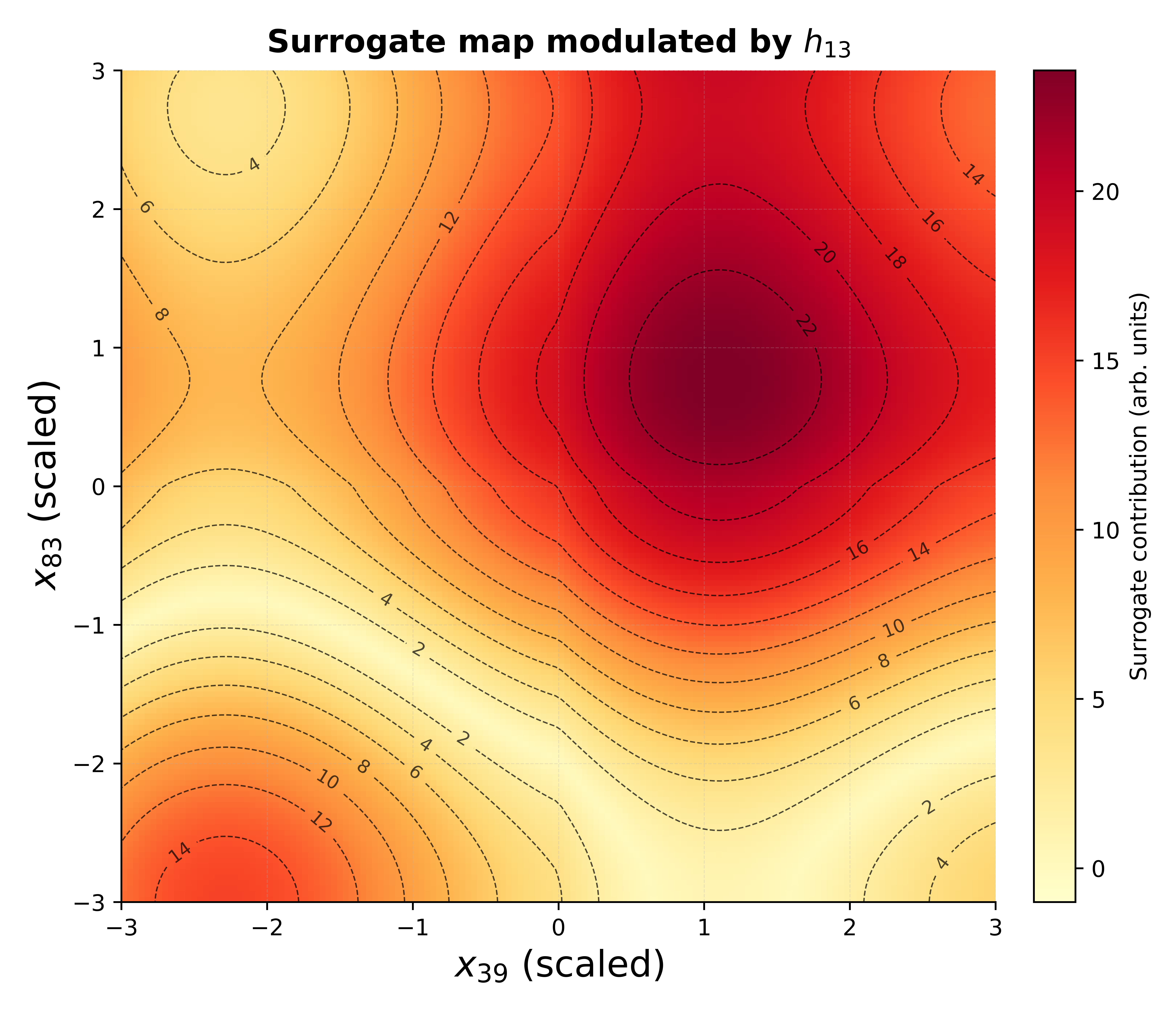} \\
        \includegraphics[width=0.4\textwidth]{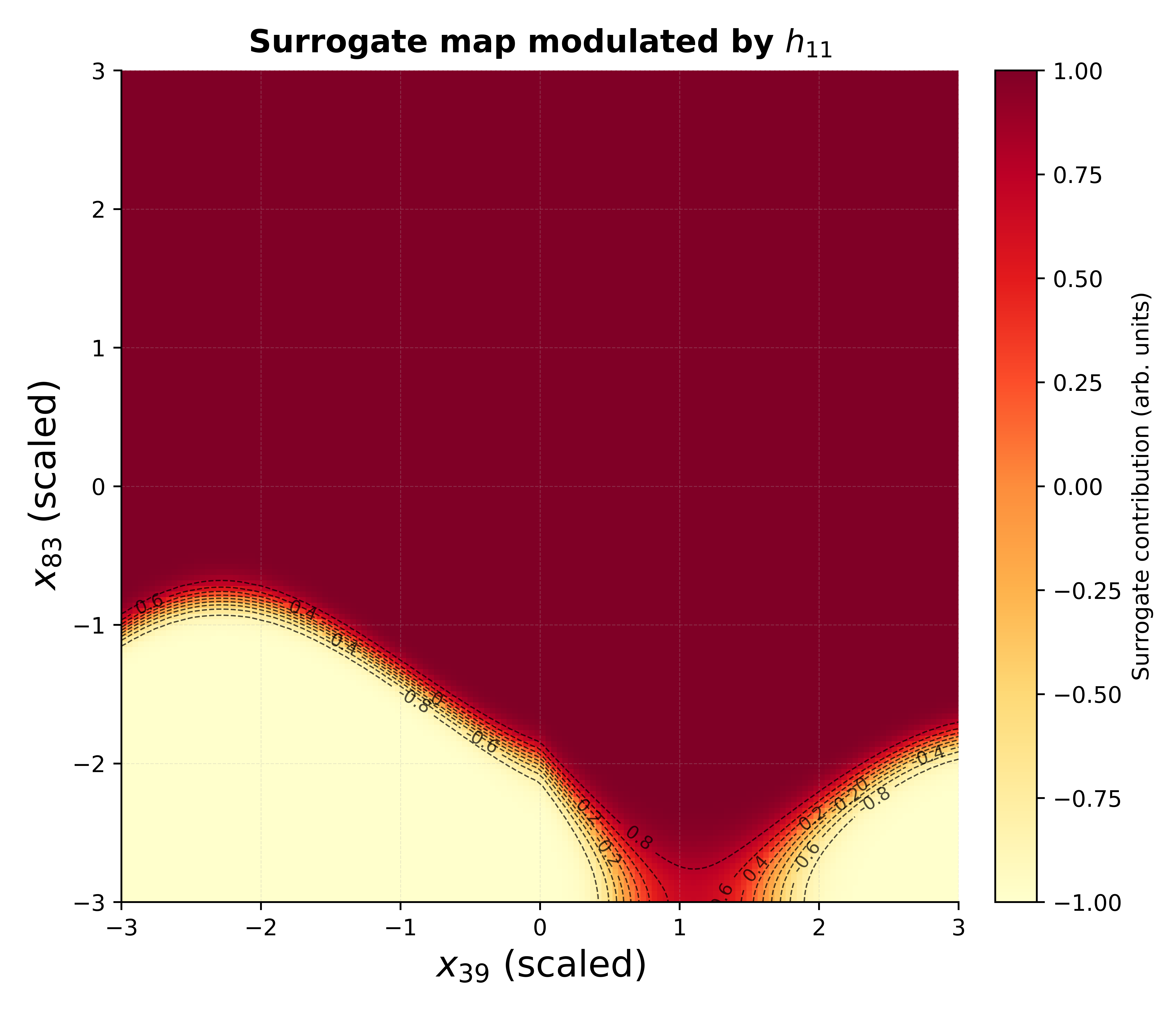} \\

\caption{Top: Seebeck feature attribution scores computed using the KAN model. The ten descriptors with the highest attribution scores were selected for interpretability analysis. Middle and bottom: Fully modulated surrogate functions for the Seebeck coefficient model, constructed using the most relevant input descriptors.}
    \label{fig:s_attr_contour}
\end{figure}

The first step in achieving symbolic interpretability is the identification of the most influential input descriptors. 
This was carried out using the KAN attribution score (see Fig.~\ref{fig:band_gap_attr_contours} and Fig.~\ref{fig:s_attr_contour}), which directly quantifies the sensitivity of the predicted property to each input. The descriptors with the highest attribution scores were selected for 
further analysis: $(x_{39},x_{68})$ for band gap and $(x_{39},x_{83})$ for the Seebeck coefficient.

\begin{figure}[htbp]
    \centering
        \includegraphics[width=0.5\textwidth]{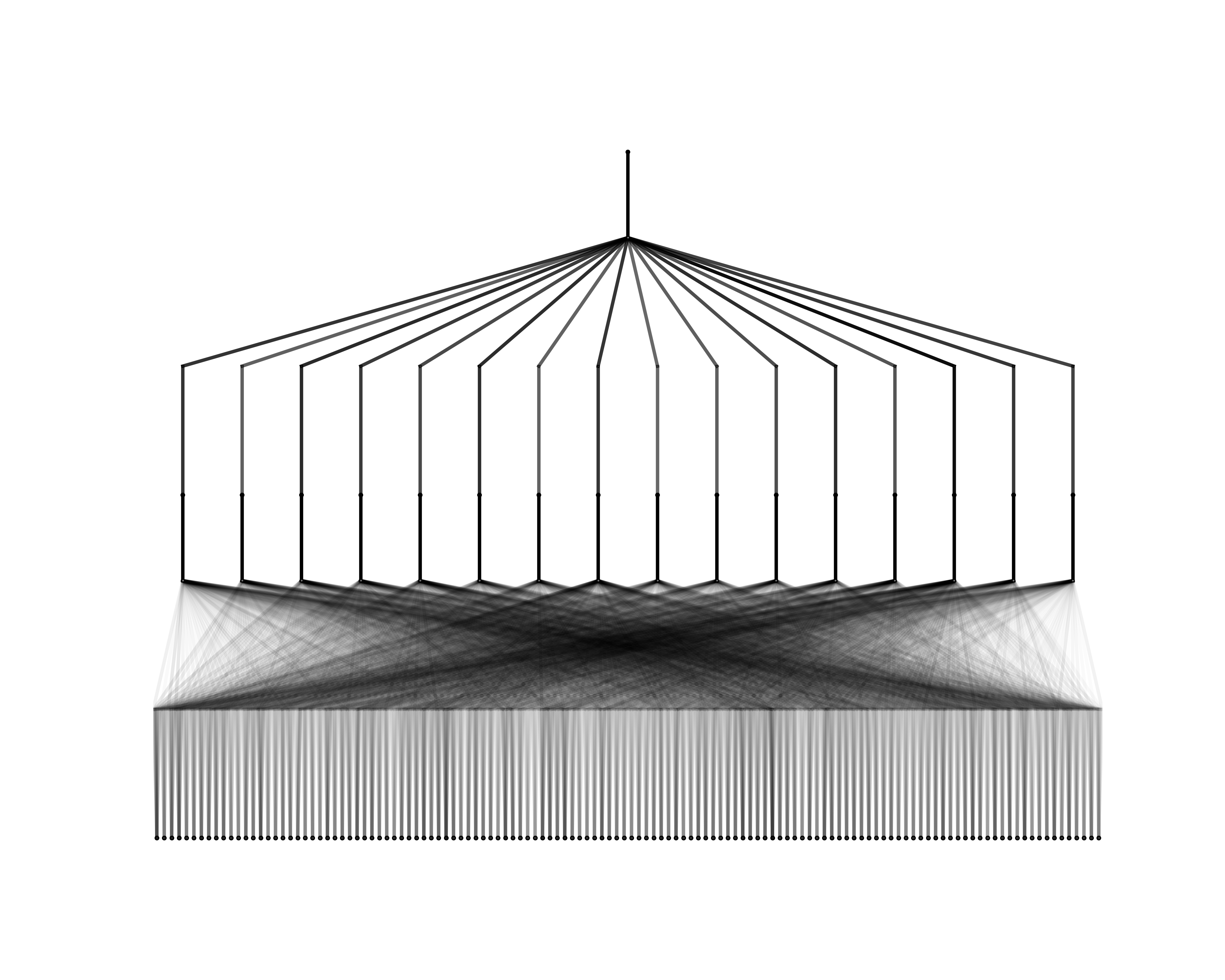} \\
    \includegraphics[width=0.5\textwidth]{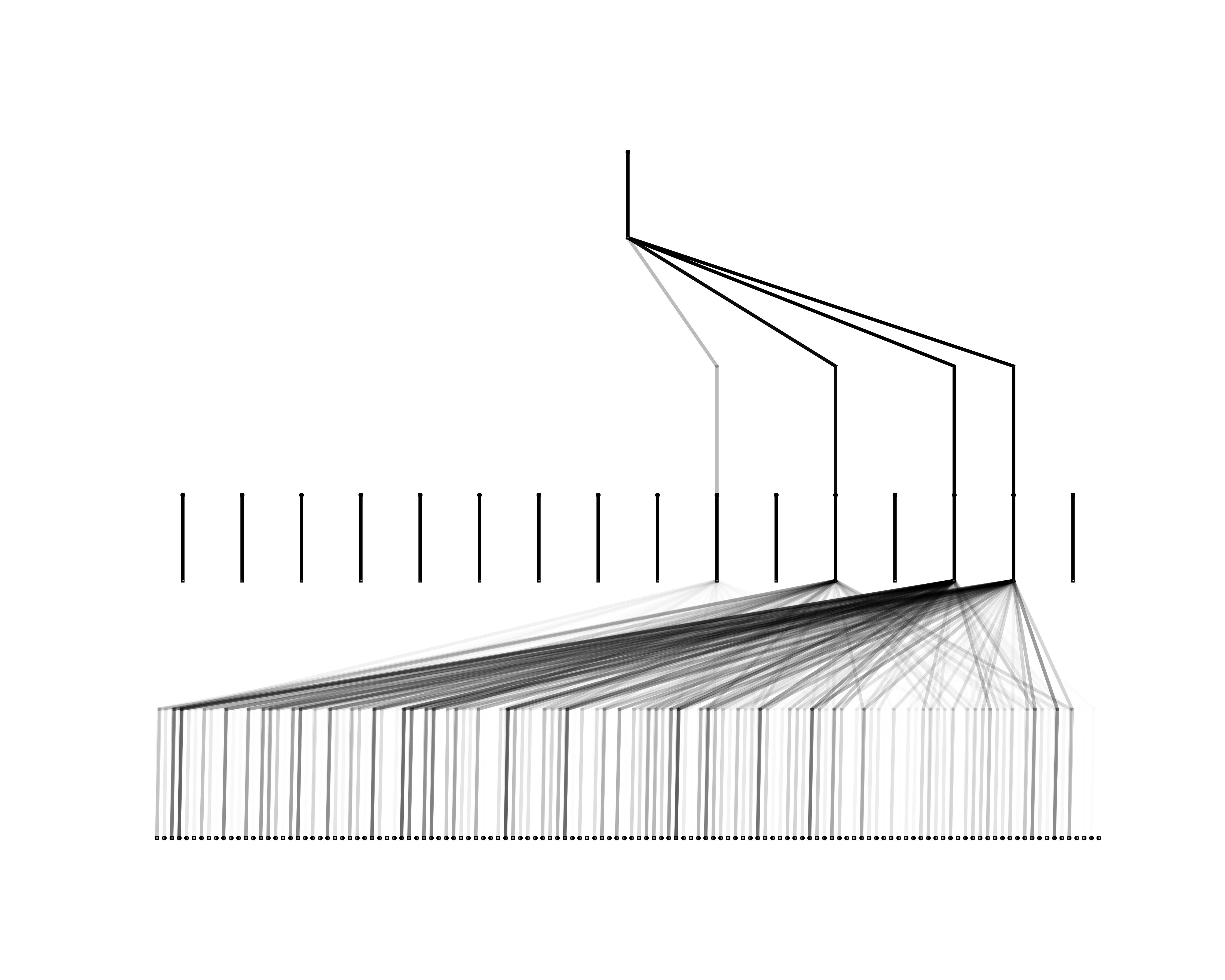}
      \includegraphics[width=0.5\textwidth]{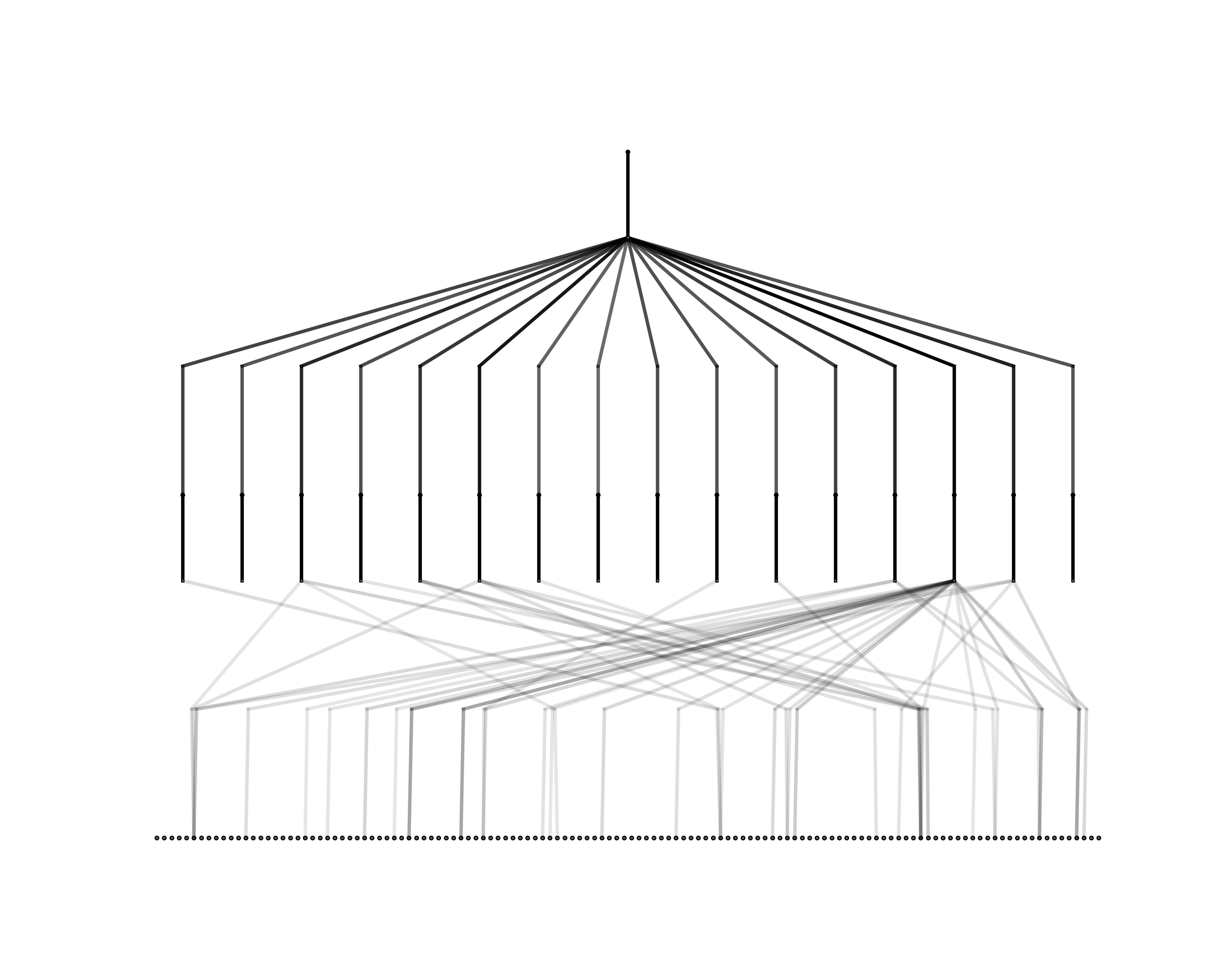}
    
\caption{Kolmogorov--Arnold Network architectures. Top: pre-training fully connected architecture. Middle: Band Gap; Bottom: Seebeck coefficient. The architectures were obtained after training to the performance levels reported in Tables~\ref{tab:kan_metrics}. 
The visualisations highlight how only a subset of edges contributes significantly to the learned structure--property mapping. 
This sparsity enables pruning of redundant connections, simplifying the network while retaining predictive accuracy and interpretability.}
    \label{fig:KAN-architectures}
\end{figure}

Redundant edges and nodes with negligible contribution were pruned. 
For each input feature, the input-hidden edge relevance was computed as the product of the spline coefficients' magnitude 
with associated scaling parameters. Node relevance was then defined as the product of the total incoming edge strength and 
the strength of the outgoing connection to the output. This definition ensures that a node is considered important only if 
it integrates significant contributions and transmits them effectively. Edges with attribution scores below 0.02 
were discarded. After pruning, the simplified models retained predictive performance within $R^2$ variations of 0.02, while revealing sparse, interpretable subnetworks (see Fig.~\ref{fig:KAN-architectures}).
The index set $\mathcal{J}$ identifies the subset of active hidden nodes retained after pruning, each contributing a simplified symbolic component $\phi_j$ to the reconstructed analytic expression.

To visualise the learned structure--property mapping, we restricted attention to the two most relevant descriptors for each target. 
The general pre-activation of a hidden node $j$ can be expressed as
\begin{equation}
S_j(x_a,x_b) \;=\; \sum_{m=1}^{16} w^{(a)}_m\, g^{(a)}_m(x_a)
\;+\; \sum_{n=1}^{16} w^{(b)}_n\, g^{(b)}_n(x_b),
\end{equation}
where $x_a,x_b$ are the two selected descriptors, $g^{(a)}_m,g^{(b)}_n$ are the symbolic edge functions, and 
$w^{(a)}_m,w^{(b)}_n$ the corresponding weights (either uniform or proportional to $R^2$).

The hidden activation is then
\begin{equation}
h_j(x_a,x_b) = \phi_j\!\big(S_j(x_a,x_b)\big),
\end{equation}
where $\phi_j$ is the symbolic hidden--output activation.  
Summing over all active hidden units $\mathcal{J}$ gives the two-descriptor surrogate for the target property:
\begin{equation}
y(x_a,x_b) \;=\; \sum_{j\in\mathcal{J}} h_j(x_a,x_b).
\end{equation}

This representation can be visualised as two-dimensional heatmaps, showing how the output varies with respect to pairs of 
descriptors while all other inputs are fixed at representative values (see Figs.~\ref{fig:band_gap_attr_contours} and \ref{fig:s_attr_contour}).

A further simplification can be achieved by retaining only the most relevant hidden nodes for each descriptor pair. 
For the band gap model, analysis revealed that $x_{39}$ and $x_{68}$ dominate through nodes $h_9$ and $h_{13}$. 
For the Seebeck coefficient, $x_{39}$ and $x_{83}$ are channelled primarily through $h_{11}$ and $h_{13}$. 
Restricting the surrogate to these nodes yields compact expressions that preserve the dominant nonlinear mechanisms.

\textbf{Band gap surrogate:}
\begin{align}
h_{9}(x_{39},x_{68}) &= \cos\!\big(S(x_{39},x_{68})\big), \\
h_{13}(x_{39},x_{68}) &= \sin\!\big(S(x_{39},x_{68})\,\big), \\
y_{\mathrm{BG}}(x_{39},x_{68}) &= h_{9}(x_{39},x_{68}) + h_{13}(x_{39},x_{68}).
\end{align}

\textbf{Seebeck surrogate:}
\begin{align}
h_{11}(x_{39},x_{83}) &= \tanh\!\big(S(x_{39},x_{83})\big), \\
h_{13}(x_{39},x_{83}) &= \big|\,S(x_{39},x_{83})\,\big|, \\
y_{\mathrm{S}}(x_{39},x_{83}) &= h_{11}(x_{39},x_{83}) + h_{13}(x_{39},x_{83}).
\end{align}

\begin{figure}[htbp]
    \centering

    \begin{tabular}{cc}
     \includegraphics[width=0.5\textwidth]{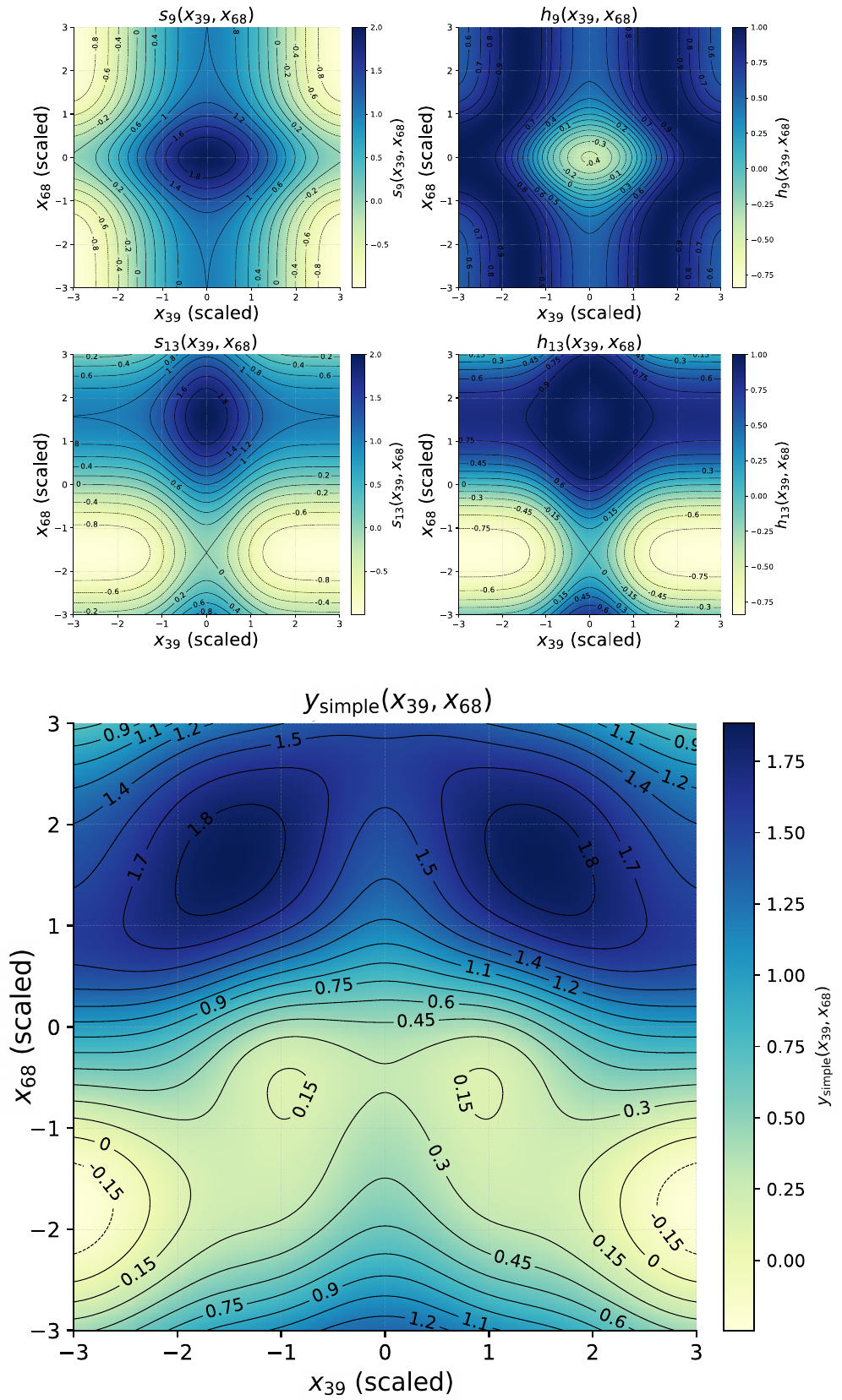}

    \end{tabular}
    \caption{Simplified surrogate maps from the most relevant hidden nodes. 
    Top: pre-activation ($s_{11}$, $s_{13}$) and post-activation ($h_{9}$, $h_{13}$). 
    Bottom: combined surrogate output $y_{\mathrm{BG}}(x_{39},x_{68})$.}
    \label{fig:b_g_simpl_contour}
\end{figure}

\begin{figure}[htbp]
    \centering
    \begin{tabular}{cc}
         \includegraphics[width=0.5\textwidth]{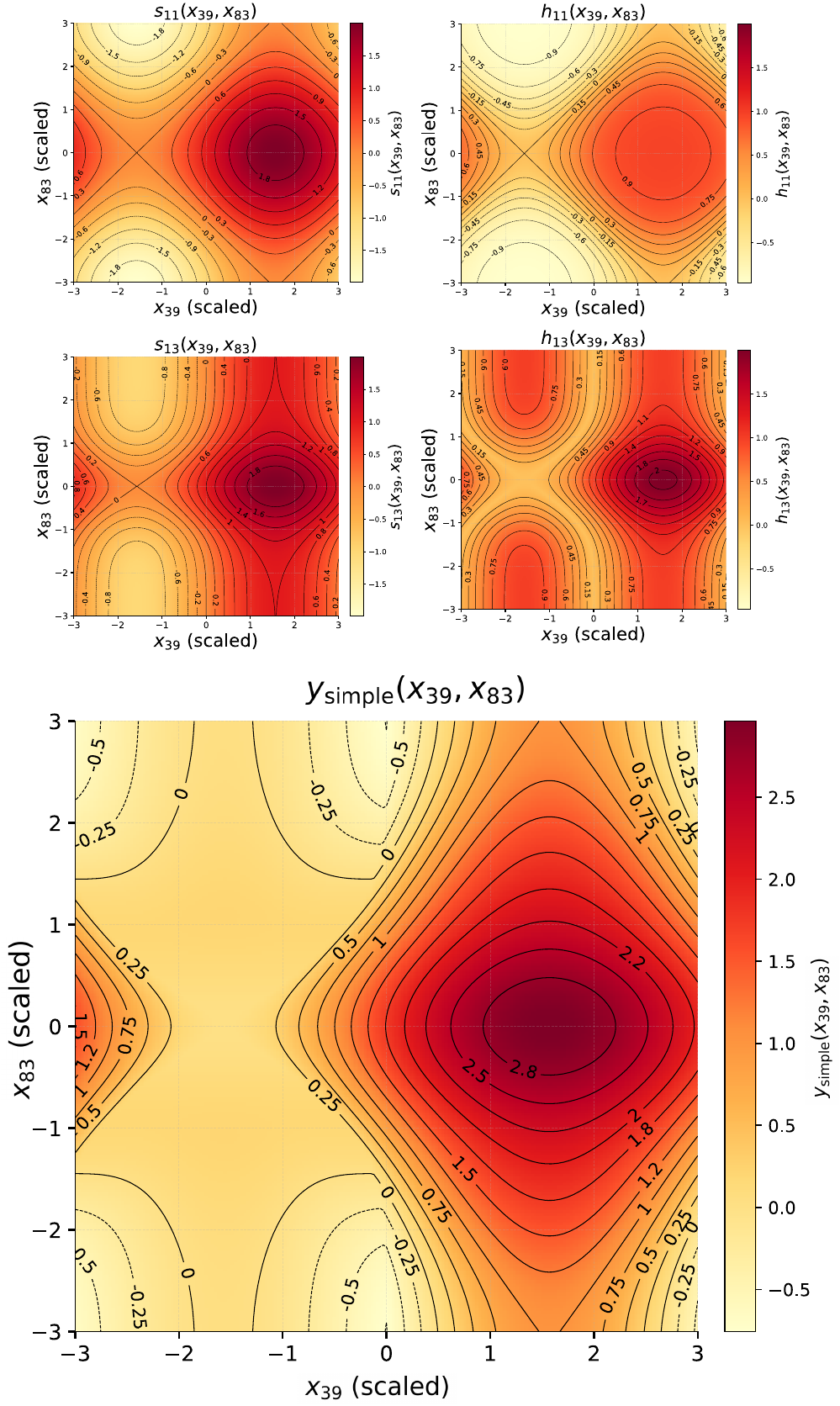}

    \end{tabular}\\[1ex]
    \caption{Simplified surrogate maps for Seebeck coefficient. Top: pre-activations ($s_{11}$, $s_{13}$) and activations ($h_{11}$, $h_{13}$). Bottom: combined surrogate output $y_{\mathrm{S}}(x_{39},x_{83})$.}
    \label{fig:seebeck-contours}
\end{figure}

These simplified surrogates were used to generate four heatmaps (two per property, one per node as shown in Figs.~\ref{fig:b_g_simpl_contour} and \ref{fig:seebeck-contours}) together with the combined output maps. 
The node-specific maps highlight distinct nonlinear responses, while the combined maps represent the net predicted property. 
Together, they provide interpretable insight into how descriptor pairs control band gap and Seebeck coefficient. 
For completeness, Table~\ref{tab:symbolic_x68_bandgap} lists the symbolic fits obtained for edges connecting  descriptor $x_{68}$ to the first hidden layer in the band gap model.

The heatmaps reveal cooperative effects between descriptors:  
for the band gap, oscillatory modulations from $x_{39}$ interact with Gaussian-like localisations from $x_{68}$, 
while for the Seebeck coefficient, trigonometric oscillations along $x_{39}$ combine with saturating and Gaussian, together with the sinusoidal oscillations responses of $x_{83}$. 
Such patterns qualitatively recover expected physical behaviour: for example, Seebeck enhancement when reduced band gap coincides with 
increased carrier concentration.
The detailed symbolic approximations for the edges involving descriptors $x_{39}$ and $x_{83}$ are summarised in Table~\ref{tab:symbolic_x38}, providing explicit forms for the Seebeck coefficient surrogate.

These results demonstrate the strength of KANs for reverse engineering in materials design:  
they expose explicit functional forms that can guide electronic and thermoelectric optimisation. 
Nevertheless, the approach has limitations: two-dimensional projections cannot capture the full high-dimensional descriptor space, 
and hidden-node surrogates may not correspond to unique physical mechanisms. 
Future work should combine descriptor dimensionality reduction with KAN symbolic extraction, 
to obtain minimal, physically meaningful descriptor sets. This would allow KANs to realise their full potential as interpretable surrogates 
for structure--property mapping in materials science.

\begin{table}[htbp]
\centering
\caption{Symbolic expressions fitted to the functions associated with the edges connecting input feature $x_{68}$ to the first hidden layer in the band gap prediction model. The coefficient of determination ($R^2$) quantifies the quality of each symbolic fit, and $c$ denotes the corresponding function complexity.}

\begin{tabular}{ccccccc}
\hline
Layer & In\_idx & Out\_idx & Function & $R^2$ & $c$ \\
\hline
0 & 68 & 0  & gaussian & 0.9165 & 3 \\
0 & 68 & 1  & abs      & 0.9297 & 3 \\
0 & 68 & 2  & abs      & 0.9535 & 3 \\
0 & 68 & 3  & sin      & 0.9923 & 2 \\
0 & 68 & 4  & gaussian & 0.9876 & 3 \\
0 & 68 & 5  & sin      & 0.9894 & 2 \\
0 & 68 & 6  & cos      & 0.9750 & 2 \\
0 & 68 & 7  & sin      & 0.9548 & 2 \\
0 & 68 & 8  & cos      & 0.9648 & 2 \\
0 & 68 & 9  & sin      & 0.9933 & 2 \\
0 & 68 & 10 & gaussian & 0.9911 & 3 \\
0 & 68 & 11 & gaussian & 0.9881 & 3 \\
0 & 68 & 12 & cos      & 0.8568 & 2 \\
0 & 68 & 13 & sin      & 0.9856 & 2 \\
0 & 68 & 14 & cos      & 0.9637 & 2 \\
0 & 68 & 15 & abs      & 0.9951 & 3 \\
\hline
\hline
0 & 39 & 0  & sin      & 0.9907 & 2 \\
0 & 39 & 1  & sin      & 0.8074 & 2 \\
0 & 39 & 2  & cos      & 0.9824 & 2 \\
0 & 39 & 3  & sin      & 0.9841 & 2 \\
0 & 39 & 4  & cos      & 0.8525 & 2 \\
0 & 39 & 5  & abs      & 0.9931 & 3 \\
0 & 39 & 6  & gaussian & 0.8167 & 3 \\
0 & 39 & 7  & abs      & 0.9664 & 3 \\
0 & 39 & 8  & cos      & 0.9866 & 2 \\
0 & 39 & 9  & gaussian & 0.9285 & 3 \\
0 & 39 & 10 & sin      & 0.9917 & 2 \\
0 & 39 & 11 & gaussian & 0.9848 & 3 \\
0 & 39 & 12 & abs      & 0.9938 & 3 \\
0 & 39 & 13 & gaussian & 0.9896 & 3 \\
0 & 39 & 14 & cos      & 0.9428 & 2 \\
0 & 39 & 15 & sin      & 0.9375 & 2 \\
\hline
\hline
1 & 0  & 0  & abs      & 0.9866 & 3 \\
1 & 1  & 0  & cos      & 0.9696 & 2 \\
1 & 2  & 0  & abs      & 0.9087 & 3 \\
1 & 3  & 0  & tanh     & 0.9957 & 3 \\
1 & 4  & 0  & gaussian & 0.9714 & 3 \\
1 & 5  & 0  & tanh     & 0.9833 & 3 \\
1 & 6  & 0  & tanh     & 0.8280 & 3 \\
1 & 7  & 0  & sin      & 0.9655 & 2 \\
1 & 8  & 0  & tanh     & 0.9961 & 3 \\
1 & 9  & 0  & cos      & 0.5921 & 2 \\
1 & 10 & 0  & gaussian & 0.9877 & 3 \\
1 & 11 & 0  & gaussian & 0.9984 & 3 \\
1 & 12 & 0  & cos      & 0.9949 & 2 \\
1 & 13 & 0  & sin      & 0.9992 & 2 \\
1 & 14 & 0  & sin      & 0.9801 & 2 \\
1 & 15 & 0  & abs      & 0.9801 & 3 \\
\hline
\end{tabular}
\label{tab:symbolic_x68_bandgap}
\end{table}

\begin{table}[htbp]
\centering
\caption{Symbolic expressions approximating the functions along edges connecting input features $x_{39}$ and $x_{83}$ to the first hidden layer in the Seebeck coefficient prediction model. The coefficient of determination ($R^2$) measures the fit accuracy, while $c$ indicates the symbolic function complexity.}
\begin{tabular}{cccccc}
\hline
Layer & In\_idx & Out\_idx & Function & $R^2$ & $c$ \\
\hline
0 & 39 & 0  & cos & 0.9860 & 2 \\
0 & 39 & 1  & sin & 0.8883 & 2 \\
0 & 39 & 2  & cos & 0.9904 & 2 \\
0 & 39 & 3  & abs & 0.9519 & 3 \\
0 & 39 & 4  & abs & 0.9419 & 3 \\
0 & 39 & 5  & $x$ & 0.9464 & 1 \\
0 & 39 & 6  & sin & 0.9880 & 2 \\
0 & 39 & 7  & cos & 0.9582 & 2 \\
0 & 39 & 8  & cos & 0.7517 & 2 \\
0 & 39 & 9  & abs & 0.9604 & 3 \\
0 & 39 & 10 & sin & 0.9631 & 2 \\
0 & 39 & 11 & sin & 0.9922 & 2 \\
0 & 39 & 12 & cos & 0.9742 & 2 \\
0 & 39 & 13 & sin & 0.9916 & 2 \\
0 & 39 & 14 & cos & 0.8838 & 2 \\
0 & 39 & 15 & cos & 0.9856 & 2 \\
\hline
\hline
0 & 83 & 0  & $x$       & 0.8660 & 1 \\
0 & 83 & 1  & $x$       & 0.8896 & 1 \\
0 & 83 & 2  & cos       & 0.9444 & 2 \\
0 & 83 & 3  & sin       & 0.9659 & 2 \\
0 & 83 & 4  & sin       & 0.9918 & 2 \\
0 & 83 & 5  & cos       & 0.9922 & 2 \\
0 & 83 & 6  & cos       & 0.9658 & 2 \\
0 & 83 & 7  & cos       & 0.9842 & 2 \\
0 & 83 & 8  & cos       & 0.8703 & 2 \\
0 & 83 & 9  & cos       & 0.9800 & 2 \\
0 & 83 & 10 & abs       & 0.7747 & 3 \\
0 & 83 & 11 & cos       & 0.9934 & 2 \\
0 & 83 & 12 & $x$       & 0.9407 & 1 \\
0 & 83 & 13 & gaussian  & 0.9729 & 3 \\
0 & 83 & 14 & abs       & 0.8528 & 3 \\
0 & 83 & 15 & cos       & 0.9660 & 2 \\
\hline
\hline
1 & 0  & 0 & tanh     & 0.8438 & 3 \\
1 & 1  & 0 & tanh     & 0.7225 & 3 \\
1 & 2  & 0 & tanh     & 0.9376 & 3 \\
1 & 3  & 0 & cos      & 0.6911 & 2 \\
1 & 4  & 0 & arctan   & 0.9858 & 4 \\
1 & 5  & 0 & tanh     & 0.9463 & 3 \\
1 & 6  & 0 & tanh     & 0.8150 & 3 \\
1 & 7  & 0 & cos      & 0.5992 & 2 \\
1 & 8  & 0 & gaussian & 0.8841 & 3 \\
1 & 9  & 0 & cos      & 0.9630 & 2 \\
1 & 10 & 0 & tanh     & 0.7758 & 3 \\
1 & 11 & 0 & tanh     & 0.9521 & 3 \\
1 & 12 & 0 & tanh     & 0.9884 & 3 \\
1 & 13 & 0 & abs      & 0.9843 & 3 \\
1 & 14 & 0 & tanh     & 0.9791 & 3 \\
1 & 15 & 0 & tanh     & 0.9377 & 3 \\
\hline
\end{tabular}
\label{tab:symbolic_x38}
\end{table}

These surrogate maps capture how pairs of descriptors cooperate to shape the prediction, revealing nonlinear interactions that are difficult to infer from attribution scores alone. While limited to two-dimensional projections, they provide interpretable insights into the structure--property relationships encoded by the KAN. Future extensions combining descriptor-reduction methods with symbolic KAN analysis may enable extraction of compact, physically meaningful descriptor sets.

\section{Discussion}

Model discovery has long been a central challenge in the physical and computational sciences. Traditional approaches have generally split into two distinct paradigms. On one hand, machine learning methods achieve impressive predictive accuracy, but typically behave as black boxes, offering limited mechanistic insight into the underlying system dynamics. On the other hand, sparse-optimization and nonlinear-dynamics approaches yield explicit, interpretable mathematical equations, but they are only applicable when the system admits an intrinsically sparse representation. Each of these paradigms thus carries significant limitations.

In this work we have shown that Kolmogorov--Arnold Networks (KANs) provide a principled bridge between these approaches. Unlike conventional neural networks, KANs retain predictive performance while also exposing how inputs influence outputs through interpretable activation functions. This allows the model not only to approximate dynamics but also to reveal symbolic surrogates of governing relationships, offering a new pathway for scientific model discovery. 

A key scientific contribution of our analysis lies in demonstrating that KANs can recover meaningful functional structures even when sparsity-based approaches fail. In functional space, there may exist infinitely many ``shadowing'' functions that reproduce the same dynamics without necessarily sharing the exact analytical form of the true governing equations. KANs naturally identify such shadowing functions, depending on architecture and regularisation, and thereby provide flexible but interpretable representations of the system’s behaviour. This suggests that KANs are not limited to symbolic regression in the narrow sense, but instead can map high-dimensional nonlinear processes into compact analytical surrogates that preserve dynamical fidelity.

Beyond their immediate predictive role, the scientific value of KANs lies in their ability to integrate domain knowledge with data-driven inference. By constraining or interpreting the learned symbolic functions in light of physical principles, one can obtain mechanistically meaningful representations that advance both understanding and control of real-world systems. 

Overall, KANs therefore provide a principled bridge between classical approximation theory and modern machine learning. They combine competitive predictive accuracy with structural transparency, enabling interpretable model discovery in contexts where neither traditional machine learning nor sparsity-optimization approaches suffice. This dual capability makes KANs especially valuable for scientific applications that demand both performance and insight, pointing toward a broader paradigm where data-driven methods contribute directly to mechanistic understanding of complex dynamical systems.

\section{Conclusion}

We have demonstrated that Kolmogorov--Arnold Networks (KANs) offer a powerful and interpretable alternative to traditional machine learning approaches for predicting key thermoelectric properties, such as the Seebeck coefficient and electronic band gap. By leveraging their functional decomposition architecture, KANs deliver predictive performance comparable to standard multilayer perceptrons while providing symbolic surrogates that reveal the underlying structure–property relationships encoded in the data.

Our comparative analysis shows that KANs maintain high accuracy across both electronic and transport properties, despite their increased computational demands. The symbolic extraction pipeline enabled us to identify the most influential descriptors, prune redundant connections, and reconstruct compact analytical expressions that qualitatively align with known physical mechanisms. In particular, we highlighted how specific combinations of input features shape model outputs through interpretable hidden activations, offering insights that are otherwise inaccessible in conventional black-box networks.

These results position KANs as a practical framework for scientific model discovery and reverse engineering in materials science. By producing explicit functional maps between structural descriptors and target properties, KANs can guide the rational design of materials with tailored thermoelectric performance. Future extensions may integrate KANs with generative models, descriptor selection schemes, or physics-informed constraints to further enhance interpretability and accelerate the discovery pipeline.

In summary, this work establishes the feasibility of KANs as both predictive tools and interpretable models for complex materials datasets, marking a step toward transparent and data-driven materials design. These findings highlight the potential of spline-based functional architectures to bridge interpretability and predictive power in materials informatics. Moreover, the demonstrated compositional invariance and stability of KANs—particularly in electronically complex systems—underscore their suitability for robust and physically consistent materials screening at scale.

\section*{Acknowledgments}

The authors thank the National Computational Infrastructure (NCI) for providing the high-performance computing resources used in this work. OI also acknowledges the support of the 3DS Science Ambassador Program.

\section*{Funding}

This work was supported by the Australian Research Council (ARC) under FL230100176. MF acknowledges support from the Australian Japan Innovation Funding (AJIF) programme. 
OI was supported by the Advanced Cyberinfrastructure Coordination Ecosystem: Services \& Support (ACCESS) program, award CHE200122, funded by NSF grants \#2138259, \#2138286, \#2138307, \#2137603, and \#2138296.

\section*{Authors' Contributions}

MF conceived and coordinated the study, performed the models training and optimisation, carried out all data analyses, and wrote the initial draft of the manuscript. 
KSN generated the descriptors dataset used for model development. 
MJF, OI, and CS contributed to scientific discussions, interpretation of the results, and critical revision of the manuscript. 
All authors reviewed and approved the final version of the manuscript.

\section*{Code Availability}

All scripts used for data preprocessing, feature integration, model training, KAN optimisation, and cross-validation are openly available at:
https://github.com/fronzi/Kolmogorov-Arnold-Networks-Thermoelecric-Materials-Design/tree/main

\section*{Competing Interests}

The authors declare no competing interests.

\bibliographystyle{rsc}
\bibliography{KANs.bib}

\end{document}